\documentclass{article}

\usepackage{arxiv}

\usepackage[utf8]{inputenc} % allow utf-8 input
\usepackage[T1]{fontenc}    % use 8-bit T1 fonts
\usepackage{hyperref}       % hyperlinks
\usepackage{url}            % simple URL typesetting
\usepackage{booktabs}       % professional-quality tables
\usepackage{amsfonts}       % blackboard math symbols
\usepackage{nicefrac}       % compact symbols for 1/2, etc.
\usepackage{microtype}      % microtypography
\usepackage{lipsum}		% Can be removed after putting your text content
\usepackage{graphicx}
\usepackage{natbib}
\usepackage{doi}
\usepackage{subfigure}
\usepackage{longtable}
\usepackage{amsmath}
\usepackage{bm}
\usepackage{multirow}
\newtheorem{Algorithm}{Algorithm}
\usepackage{makecell}
\newtheorem{theorem}{Theorem}
\newtheorem{Example}{Example}
\title{Efficient Tests for Testing in Two-way ANOVA under Heteroscedasticity}

\author{ Anjana Mondal \\
	Department of Mathematics and Statistics\\
	Indian Institute of Technology Tirupati\\
	Tirupati, 517619 \\
	\texttt{anjanam@iittp.ac.in} \\
	\And
	Somesh Kumar \\
	Department of Mathematics\\
	Indian Institute of Technology Kharagpur\\
	Kharagpur, 721302 \\
	\texttt{smsh@maths.iitkgp.ac.in} \\
}

\renewcommand{\shorttitle}{\textit{arXiv} Template}

\hypersetup{
pdftitle={A template for the arxiv style},
pdfsubject={q-bio.NC, q-bio.QM},
pdfauthor={David S.~Hippocampus, Elias D.~Striatum},
pdfkeywords={First keyword, Second keyword, More},
}

\begin{document}
\maketitle

\begin{abstract}
New tests are developed for two-way ANOVA models with heterogeneous error variances. The testing problems are considered for testing the significant interaction effects, simple effects, and treatment effects. The likelihood ratio tests (LRTs) and simultaneous comparison tests are derived for all three problems. Hill climbing algorithms have been proposed to compute the maximum likelihood estimators (MLEs) of parameters under the restrictions on the null and alternative hypotheses. It is proved that the proposed algorithms converge to the MLEs. A parametric bootstrap algorithm is provided for the computation of the critical points. The simulated power values of the proposed tests are compared with two existing tests. For testing main effects in the additive ANOVA model, the LRT appears to be about  $30\%$ to $50\%$ gain in power over the available tests. Also, the proposed tests for the interaction and simple effects are seen to have comparable power and size performance to the existing tests. The behavior of the proposed tests under the non-normal error distribution is also discussed. Four real data sets are used to demonstrate the application of the proposed tests. A software package is made in `R' to make it simple to apply the tests to experimental data sets.
\end{abstract}

% keywords can be removed

\section{Introduction}
\begin{figure}[h!]
\centering
\includegraphics[width=0.90\linewidth]{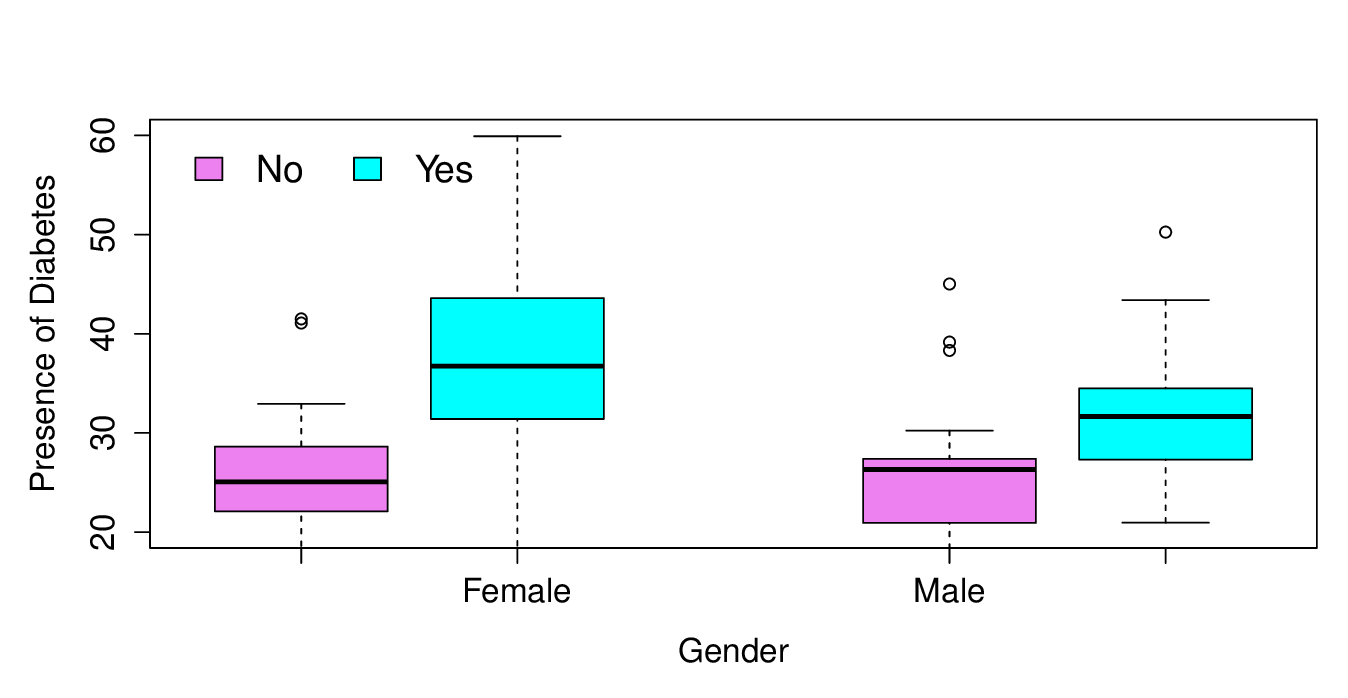}
\caption{Box plots of BMI values according to gender and status of diabetes}
\label{box plot}
\end{figure}

Many research in biology, agriculture, clinical trials, genetics, and industrial design to compare treatment effects using two-way ANOVA models.  A two-way ANOVA model can appear in a variety of situations. A diabetes prediction dataset has been picked as an inspiring example from
\url{https://www.kaggle.com/datasets/iammustafatz/diabetes-prediction-dataset }. Here, we look into how a person's gender and diabetes status affect their body mass index (BMI). A two-way ANOVA is used to model the dataset, with factors $A$ denoting "presence of diabetes" with two levels (whether diabetes is present or not) and $B$ denoting gender with two levels (male and female). The BMI levels of subjects are used as response variables. Box plots of BMI data for male and female individuals with and without diabetes are shown in Figure \ref{box plot}.  This implies that the response variable is influenced by both factors.

Conventional ANOVA assumes that error variances are homogeneous and uses traditional $F$-tests. This assumption might not hold true in a lot of real-world scenarios. This is due to differences in experimental conditions, changes in operators, variations due to the sampling scheme, etc. Applying the traditional $F$-tests in these situations could result in an incorrect interpretation of the data; see, for example, Bishop and Dudewicz (1981),  Hsuing {\it{et al}.}. (1996), Ananda and Weerahandi (1997), and  Zhang (2012). Two-way ANOVA with heterogeneity has been the subject of a few studies.

A two-step test strategy for $r$-way ANOVA with heteroscedasticity was proposed by Bishop and Dudewicz (1981). Ananda and Weerahandi (1997) designed a generalized $p$-value approach to determine the significance of treatment effects and interaction under heteroscedasticity.  Using an actual data set, they have demonstrated that the null hypothesis is not rejected by the traditional $F$-test, even if the data appears to offer sufficient evidence to do so.  Simulations indicate that their tests beat traditional $F$-tests in terms of controlling the type-I error rates and attaining good powers. Bao and Ananda (2001) used simulations to evaluate the size and power of the traditional $F$-test, and the modified $F$-test of Ananda and Weerahandi (1997) in order to assess the importance of the interaction effects. According to the reported results, the generalized $F$-test is better than the traditional $F$-test when dealing with heterogeneous error variances. When there are several levels of one or both factors, Wang and Akritas (2006) have suggested some asymptotic tests for testing interaction and treatment effects under heteroscedasticity and non-normality. Alin and Kurt (2006) have provided a test for the importance of interaction effects for data with a single observation per cell. An approximate $F$-test for testing in a two-way crossed model was proposed by Zhang (2012).  Here, the provided data must be used to estimate the degree of freedom. A simulation study demonstrates that under heteroscedasticity, the traditional $F$-test exhibits overestimated type-I errors. Parametric bootstrap tests have been developed by Xu {\it{et al.}} (2013) for the main effects and interaction significance in a two-way crossed model with heterogeneous error variances. To test in non-parametric heteroscedastic factorial designs, Brunner {\it{et al.}} (1997) provided a few approximate tests. Pauly {\it{et al.}}. (2015) have proposed some asymptotic and permutation tests in general factorial designs. Parametric bootstrap tests have been developed by Xu {\it{et al}.} (2015) in an additive model.  In a two-way crossed model, Zhang (2015) provided simultaneous confidence intervals for pairwise multiple comparisons of the treatment effects. Ananda {\it{et al.}} (2023) introduced two hypothesis tests for the two-way heteroscedastic ANOVA model using the generalized $p$-value framework. They showed that one of the proposed tests is numerically equivalent to the parametric bootstrap (PB) test of Xu {\it{et al.}} (2013). In addition, they highlighted several advantages of the generalized p-value approach over the PB method and provided an R package, `twowaytests', on CRAN to facilitate practical implementation of their procedures.
%Although the generalized $p$-value approach offers certain advantages over the parametric bootstrap method in some settings, it's primary challenge lies in constructing an appropriate pivotal test statistic.

As a result, it should be mentioned that while tests for the two-way ANOVA model with heterogeneous error variances are available, the likelihood ratio test (LRT) has not yet been taken into consideration. In this paper, we take up this problem in a comprehensive way. The three testing problems, that of the significance of interactions, simple, and main effects are investigated. The LRT and a multiple comparisons test are developed for all three testing problems. Moreover, these procedures are seen to be either better or at least as good as the existing tests with respect to empirical size and power values.

Consider the two-way ANOVA with $a$ levels of factor $A$ and $b$ levels of factor $B$.  The observations $Y_{ijk}$ of this =can be modeled as
\begin{equation}
Y_{ijk}=\mu+\alpha_i+\beta_j+\gamma_{ij}+\epsilon_{ijk};~k=1,\ldots ,n_{ij}; i=1,\ldots ,a;j=1,\ldots ,b.
\end{equation}
Here, the overall mean is $\mu$, the effects of the $i$-th level of factor $A$ and the $j$-th level of factor $B$ are indicated by $\alpha_i$ and $\beta_j$, respectively; $\gamma_{ij}$ indicates the interaction effect caused by the $i$-th level of factor $A$ and $j$-th level of factor $B$; and $\epsilon_{ijk}$'s represent errors.
We consider the errors $\epsilon_{ijk}$ to be independent random variables which have distribution $N(0,\sigma_{ij}^2)$ for $k=1,\ldots,n_{ij};i=1,\ldots,a;j=1,\ldots,b.$ The variances $\sigma_{ij}^2$ are assumed to be unknown and unequal. The following zero-sum requirements are assumed for the estimability of parameters:
\begin{equation}\label{const}
\sum\limits_{i=1}^a\alpha_i=0,~\sum\limits_{j=1}^b\beta_j=0,~\sum\limits_{i=1}^a\gamma_{ij}=0,~\sum\limits_{j=1}^b \gamma_{ij}=0.
\end{equation}
Here, the goal is to test the hypothesis of no interaction effect; that is,
\begin{equation}\label{hp1}
 H_{0AB}:\gamma_{11}=\cdots =\gamma_{1b}=\cdots=\gamma_{a1}=\cdots =\gamma_{ab}=0~\text{against}~H_{1AB}:\gamma_{ij}\neq 0~\text{for at least one pair}~(i,j). 
\end{equation}
When the interaction effects are significant, one would like to test the simple effects for factor $A$, i.e.,
\begin{equation}\label{hp2}
H_{0A}^*:\alpha_i+\gamma_{ij}=0;~i=1,\ldots ,a;j=1,\ldots ,b~\text{against}~H_{1A}^*:\alpha_i+\gamma_{ij}\neq 0~\text{for at least one pair }~(i,j) .
\end{equation}
When there are no interaction effects, one is interested in testing the homogeneity of main effects, that is,
\begin{equation}\label{hp2}
H_{0A}:\alpha_1=\cdots =\alpha_a=0~\text{against}~H_{1A}:\alpha_{i_1}\neq\alpha_{i_2}~\text{for at least one pair}~(i_1,i_2).
\end{equation}

In this paper, we try to develop the LRT.  It is also suggested to test $H_{0A}$ against $H_{1A}$ using a different test that compares treatment effects utilizing multiple comparisons.  It is implemented using a parametric bootstrap.  For the aforementioned hypotheses, Xu {\it{et al.}} (2013) and (2015) have provided tests. They provided test statistics based on the standardized sum of squares for the known variances.  Estimators of the variances were used in place of the unknown variances.  Critical points were obtained using a parametric bootstrap (PB). The comparison of the empirical size and power values of the proposed tests to those of Xu {\it{et al.}} (2013) and (2015) has been done. With a 30–50$\%$ increase in power, the simulated results demonstrate that the LRT and MCT are more potent than Xu et al.'s tests. Xu {\it{et al}}. (2015) considered the crossed effects model. We have also calculated the power and size values of their test for simple effects, assuming no interaction. Simulation studies show that the proposed LRTs and available tests have similar power performance. The LRT provides size values that are nearly at the nominal level and does not experience any power loss. When size values are achieved for moderate and large samples, the asymptotic LRT performs better than all other tests. While certain tests for two-way ANOVA with heteroscedasticity already exist, determining their critical values often poses computational challenges. To address this, we developed the R package `TwowayANOVATests', which enables convenient application of the proposed methods to real data sets. This ease of implementation represents an added advantage of our approach over existing tests.

This article is organized as follows. In Section \ref{sec2}, tests are proposed for testing $H_{0AB}$ against $H_{1AB}$. In Section \ref{sec2.2}, the maximum likelihood estimators (MLEs)  of parameters are obtained. Section \ref{sec2.3} describes a parametric bootstrap algorithm for implementing the LRT. The multiple comparison test (MCT) has been developed in Section \ref{sec2.4}. The LRT and MCT are developed for testing simple effects in Section \ref{sec3}. The LRT, MCT, and asymptotic MCT are proposed for testing treatment effects in the additive model in Section \ref{sec4}.
%Section \ref{sec3.1} provides asymptotic tests using the null distribution of
%the test statistic from Section \ref{sec3}.
The power comparisons of the proposed tests with the available tests of Xu {\it{et al.}}. (2013) and (2015) are done in Section \ref{sec5}. Under several non-normal distributions, the size and power values of the proposed tests are presented in Section \ref{sec6}. The proposed tests are illustrated through four real data examples in Section \ref{sec7}. In Section \ref{sec8}, the developed `R' package is described in detail.

\section{Testing for Interaction Effects}\label{sec2}
In this section, the LRT is proposed for testing $H_{0AB}$ against the natural alternative hypothesis $H_{1AB}$.
We use re-parameterization as $\zeta_j=\mu+\beta_j,$ $ j=1,\ldots ,b.$ This helps in incorporating some of the zero-sum conditions of the parameters.
Let $\Omega_{0AB}$ and $\Omega$ denote the parameter spaces under $H_{0AB}$ and $H_{0AB}\cup H_{1AB}.$
These are defined below:
$$\Omega_{0AB}=\left\lbrace\left(\alpha_1,\ldots,\alpha_a,\zeta_1,\ldots ,\zeta_b,\sigma_{11}^2,\ldots ,\sigma_{ab}^2\right):\alpha_i,\zeta_j\in\mathbb{R};\sigma_{ij}\in\mathbb{R}^{+};\sum\limits_{i=1}^a\alpha_i=0\right\rbrace$$
and 
$$\Omega=\left\lbrace\left(\zeta_1,\ldots ,\zeta_b,\gamma_{11},\ldots ,\gamma_{ab},\sigma_{11}^2,\ldots ,\sigma_{ab}^2\right):\zeta_j,\gamma_{ij}\in\mathbb{R};\sigma_{ij}\in\mathbb{R}^{+};\sum\limits_{i=1}^a\gamma_{ij}=0,\sum\limits_{j=1}^b\gamma_{ij}=0\right\rbrace.$$
%To get the likelihood ratio test statistic, one needs the maximum likelihood estimators of parameters under the parameter spaces $\Omega$ and $\Omega_{0AB}$.
To get the likelihood ratio test statistic, the procedures for evaluating the MLEs of parameters under $\Omega$ and  $\Omega_{0AB}$ are discussed in the following subsections. 

\subsection{The MLEs under $\Omega$}\label{sec2.1}

Define $\bar{Y}_{ij}=\frac{1}{n_{ij}}\sum\limits_{k=1}^{n_{ij}} Y_{ijk}$, $\bar{Y}_{i.}=\frac{1}{b}\sum\limits_{j=1}^b \bar{Y}_{ij}$, $\bar{Y}_{.j}=\frac{1}{a}\sum\limits_{i=1}^a \bar{Y}_{ij}$, $\bar{Y}=\frac{1}{ab}\sum\limits_{i=1}^a\sum\limits_{j=1}^b \bar{Y}_{ij}$ and $S_{ij}^2=\frac{1}{n_{ij}-1}\sum\limits_{k=1}^{n_{ij}}(Y_{ijk}-\bar{Y}_{ij})^2$ for $i=1,\ldots ,a;j=1,\ldots ,b.$ 
In this case, the MLEs of $\mu_{ij}=\mu+\alpha_i+\beta_j+\gamma_{ij}$ and $\sigma_{ij}^2$ are respectively $\hat{\mu}_{ij\Omega}=\bar{Y}_{ij}$ and $\hat{\sigma}_{ij\Omega}^2=\frac{1}{n_{ij}}\sum\limits_{k=1}^{n_{ij}}(Y_{ijk}-\bar{Y}_{ij})^2$ for $i=1,\ldots ,a;j=1,\ldots ,b.$ Since the parameters $\mu,\alpha_i,\beta_j,\gamma_{ij}$ are the linear transforms of the vector of cell means $\mu_{ij}$'s, the MLEs are their usual unbiased estimators. That is, $\hat{\mu}_{\Omega}=\bar{Y}$, $\hat{\alpha}_{i\Omega}=\bar{Y}_{i.}-\bar{Y}$, $\hat{\beta}_{j\Omega}=\bar{Y}_{.j}-\bar{Y}$ and $\hat{\gamma}_{ij\Omega}=\bar{Y}_{ij}-\bar{Y}_{i.}-\bar{Y}_{.j}+\bar{Y}$ for $i=1,\ldots ,a;j=1,\ldots ,b.$

\subsection{The MLEs under $\Omega_{0AB}$}\label{sec2.2}

We present a method in this section for determining the MLEs of parameters under $\Omega_{0AB}$. The MLEs of parameters must meet the likelihood equations since the likelihood function is a differentiable function of parameters. The algorithm is designed to solve likelihood equations. 
With the zero-sum constraint $\sum\limits_{i=1}^a\alpha_i=0$ included, we get the log-likelihood function under $\Omega_{0AB}$ as
\begin{equation}
l_{\Omega_{0AB}}\left(\bm{\zeta},\bm{\alpha},\bm{\sigma}^2\right)=-\sum\limits_{i=1}^a \sum\limits_{j=1}^b \frac{n_{ij}}{2} \ln \sigma_{ij}^2-\sum\limits_{i=1}^{a-1}\sum\limits_{j=1}^{b} \sum\limits_{k=1}^{n_{ij}}\frac{1}{2\sigma_{ij}^2}\left(Y_{ijk}-\alpha_i-\zeta_j\right)^2 -\sum\limits_{j=1}^b\sum\limits_{k=1}^{n_{aj}}\frac{1}{\sigma_{aj}^2}\left(Y_{ajk}+\sum\limits_{i_1=1}^{a-1}\alpha_{i_1}-\zeta_j\right)^2+C,
\end{equation}
where $C$ is free from parameters.
The MLEs of $\alpha_i$, $\zeta_j$, and $\sigma_{ij}^2$ for $i=1,\ldots,a;j=1,\ldots,b.$ are represented by $\hat{\alpha}_{i\Omega_{0AB}}$, $\hat{\zeta}_{j\Omega_{0AB}}$, and $\hat{\sigma}_{ij\Omega_{0AB}}^2$, respectively.
The likelihood equations are then presented in their simplified form by
\begin{equation}\label{lk1}
\left(\hat{w}_{i\Omega_{0AB}}+\hat{w}_{a\Omega_{0AB}}\right)\hat{\alpha}_{i\Omega_{0AB}}+\hat{w}_{a\Omega_{0AB}}\sum\limits_{i_1=1;i_1\neq i}^{a-1}\hat{\alpha}_{i_1\Omega_{0AB}}=\sum\limits_{j=1}^b\hat{u}_{ij\Omega_{0AB}}\left(\bar{Y}_{ij}-\hat{\zeta}_{j\Omega_{0AB}}\right)-\sum\limits_{j=1}^b \hat{u}_{aj\Omega_{0AB}}\left(\bar{Y}_{aj}-\hat{\zeta}_{j\Omega_{0AB}}\right);~i=1,\ldots,a-1,
\end{equation}
\begin{equation}\label{lk2}
\hat{\zeta}_{j\Omega_{0AB}}=\frac{\sum\limits_{i=1}^a\hat{u}_{ij\Omega_{0AB}}\left(\bar{Y}_{ij}-\hat{\alpha}_{i\Omega_{0AB}}\right)}{\sum\limits_{i=1}^a \hat{u}_{ij\Omega_{0AB}}};~j=1,\ldots ,b
\end{equation}
and
\begin{equation}\label{lk3}
\hat{\sigma}_{ij\Omega_{0AB}}^2=\frac{1}{n_{ij}}\sum\limits_{k=1}^{n_{ij}}\left(Y_{ijk}-\hat{\alpha}_{i\Omega_{0AB}}-\hat{\zeta}_{j\Omega_{0AB}}\right)^2;~i=1,\ldots ,a;j=1,\ldots ,b,
\end{equation}
where $\hat{u}_{ij\Omega_{0AB}}=\frac{n_{ij}}{\hat{\sigma}_{ij\Omega_{0AB}}^2}$, 
$\hat{w}_{i\Omega_{0AB}}=\sum\limits_{j=1}^b \hat{u}_{ij\Omega_{0AB}},$ $\bar{Y}_{ij}=\frac{1}{n_{ij}}\sum\limits_{k=1}^{n_{ij}}Y_{ijk}$ for $i=1,\ldots ,a;j=1,\ldots ,b.$

Define 
$$W_{\Omega_{0AB}}=\text{diag}\left(\hat{w}_{1\Omega_{0AB}},\ldots ,\hat{w}_{(a-1),\Omega_{0AB}}\right)+\hat{w}_{a\Omega_{0AB}}\bm{1}_{a-1}\bm{1}_{a-1}^T,$$ $$U_{\Omega_{0AB}}=\left( u_{1\Omega_{0AB}},\ldots ,u_{(a-1)\Omega_{0AB}}\right)^T,u_{i\Omega_{0AB}}=\sum\limits_{j=1}^b\hat{u}_{ij\Omega_{0AB}}\left(\bar{Y}_{ij}-\hat{\zeta}_{j\Omega_{0AB}}\right)-\sum\limits_{j=1}^b\hat{u}_{aj\Omega_{0AB}}\left(\bar{Y}_{aj}-\hat{\zeta}_{j\Omega_{0AB}}\right);~i=1,\ldots ,a-1,$$ with $\bm{1}_{a-1}$ being the column vector of 1s of order $a-1$.
Then the matrix representation of the set of equations (\ref{lk1}) is as follows
$$W_{\Omega_{0AB}}\bm{\hat{\alpha}}_{*\Omega_{0AB}}=U_{\Omega_{0AB}},$$
where $\bm{\hat{\alpha}}_{*\Omega_{0AB}}=\left(\hat{\alpha}_{1\Omega_{0AB}},\ldots ,\hat{\alpha}_{(a-1)\Omega_{0AB}}\right)^T.$

The solutions to the aforementioned system of non-linear equations (\ref{lk1}), (\ref{lk2}), and (\ref{lk3}) can be obtained iteratively using the process described below. We have demonstrated that this technique converges to true MLEs under $\Omega_{0AB}$ in the Appendix.

%Define $\bar{Y}_{i.}=\frac{1}{b}\sum\limits_{j=1}^b\bar{Y}_{ij}$, $\bar{Y}_{.j}=\frac{1}{a}\sum\limits_{i=1}^a\bar{Y}_{ij}$ for $i=1,\ldots ,a;~j=1,\ldots ,b.$
The notations $\alpha_{i\Omega_{0AB}}^{(m)}$, $\zeta_{j\Omega_{0AB}}^{(m)}$ and $\sigma_{ij\Omega_{0AB}}^{2(m)}$ in the following algorithm represent the values of $\alpha_i$, $\zeta_j$ and $\sigma_{ij}^2$ at the $m$-th iteration for $i=1,\ldots ,a;j=1,\ldots ,b.$ 

%In this section, one iterative algorithm is given for finding the MLEs of parameters under $\Omega_{0AB}.$ Note that the MLEs of parameters must satisfy the likelihood equations. The following algorithm is obtained based on a similar technique to the earlier section.

%Define $\bar{Y}_{i.}=\frac{1}{b}\sum\limits_{j=1}^b\bar{Y}_{ij}$, $\bar{Y}_{.j}=\frac{1}{a}\sum\limits_{i=1}^a\bar{Y}_{ij}$ for $i=1,\ldots ,a;~j=1,\ldots ,b.$
%In the following algorithm, $\alpha_{i\Omega_{0AB}}^{(m)}$, $\zeta_{j\Omega_{0AB}}^{(m)}$ and $\sigma_{ij\Omega_{0AB}}^{2(m)}$ denote the $m$-th iterative values of $\alpha_i$, $\zeta_j$ and $\sigma_{ij}^2$ for $i=1,\ldots ,a;j=1,\ldots ,b.$ 
%\newpage
\begin{Algorithm}\label{algo2}
\hspace{1cm}
\begin{itemize}
\item[ Step (0,0):] Choose the starting values of the parameters as follows: 
$\alpha_{i\Omega_{0AB}}^{(0)}=\bar{Y}_{i.}$, $\zeta_{j\Omega_{0AB}}^{(0)}=\bar{Y}_{.j}-\bar{Y}$ and $\sigma_{ij\Omega_{0AB}}^{2(0)}=S_{ij}^2=\frac{1}{n_{ij}}\sum\limits_{k=1}^{n_{ij}}\left(Y_{ijk}-\bar{Y}_{ij}\right)^2;~i=1,\ldots ,a;j=1,\ldots ,b.$

\item[Step ($m$,1):] Define $u_{ij\Omega_{0AB}}^{(m-1)}=\frac{n_{ij}}{\sigma_{ij\Omega_{0AB}}^{2(m-1)}}$, 
 $w_{i\Omega_{0AB}}^{(m-1)}=\sum\limits_{j=1}^bu_{ij\Omega_{0AB}}^{(m-1)}$, $u_{i\Omega_{0AB}}^{(m-1)}=\sum\limits_{j=1}^b u_{ij\Omega_{0AB}}^{(m-1)}\left(\bar{Y}_{ij}-\zeta_{j\Omega_{0AB}}^{(m-1)}\right)-\sum\limits_{j=1}^b u_{aj\Omega_{0AB}}^{(m-1)}\left(\bar{Y}_{aj}-\zeta_{j\Omega_{0AB}}^{(m-1)}\right)$ for $i=1,\ldots ,(a-1),$ $W_{\Omega_{0AB}}^{(m-1)}=\text{diag}\left(w_{1\Omega_{0AB}}^{(m-1)},\ldots ,w_{(a-1)\Omega_{0AB}}^{(m-1)}\right)+w_{a\Omega_{0AB}}^{(m-1)}\bm{1}_{a-1}\bm{1}_{a-1}^T$, $U_{\Omega_{0AB}}^{(m-1)}=\left(u_{1\Omega_{0AB}}^{(m-1)},\ldots ,u_{(a-1)\Omega_{0AB}}^{(m-1)}\right)^T$. The value of $\bm{\alpha}_{*}=\left(\alpha_1,\ldots,\alpha_{a-1}\right)^T$ is then determined as follows $\bm{\alpha}_{*}^{(m)}=W_{\Omega_{0AB}}^{-1(m-1)}U_{\Omega_{0AB}}^{(m-1)},$ where \\$\bm{\alpha}_{*}^{(m)}=\left(\alpha_{1\Omega_{0AB}}^{(m)},\ldots ,\alpha_{(a-1)\Omega_{0AB}}^{(m)}\right)^T$  and evaluate the $m$-th iterative value of $\alpha_a$ as 
$\alpha_{a\Omega_{0AB}}^{(m)}=-\sum\limits_{i=1}^{a-1}\alpha_{i\Omega_{0AB}}^{(m)}$.
\item[Step ($m$,2):] Compute the $m$-th iterative value of $\zeta_j$ as
$$\zeta_{j\Omega_{0AB}}^{(m)}=\frac{\sum\limits_{i=1}^a u_{ij\Omega_{0AB}}^{(m-1)}\left(\bar{Y}_{ij}-\alpha_{i\Omega_{0AB}}^{(m)}\right)}{\sum\limits_{i=1}^a {u}_{ij\Omega_{0AB}}^{(m-1)}}~\text{for}~j=1,\ldots ,b.$$
\item[Step ($m$,3):] 
Utilizing the values computed in Steps $(m,1)$ and $(m,2)$, determine the $m$-th iterative value of $\sigma_{ij}^2$ as
$$\sigma_{ij\Omega_{0AB}}^{2(m)}=\frac{n_{ij}-1}{n_{ij}}S_{ij}^2+\left(\bar{Y}_{ij}-\alpha_{i\Omega_{0AB}}^{(m)}-\zeta_{j\Omega_{0AB}}^{(m)}\right)^2;~i=1,\ldots ,a;j=1,\ldots ,b.$$
\item[Step ($m$,4):] Repeat Steps ($m$, 1), ($m$, 2) and ($m$, 3). 
Terminate the algorithm if $\max\limits_{1\leq i\leq a}\mid \alpha_{i\Omega_{0AB}}^{(m-1)} -\alpha_{i\Omega_{0AB}}^{(m)}\mid\leq \epsilon$ and $\max\limits_{1\leq j\leq b}\mid \zeta_{j\Omega_{0AB}}^{(m-1)} -\zeta_{j\Omega_{0AB}}^{(m)}\mid\leq \epsilon$, where $\epsilon$ denotes the prescribed tolerance level.
\end{itemize}
%\caption{Algorithm for finding MLEs of parameters under $\Omega$}
\end{Algorithm}
The procedure starts from an initial value and iterates until successive estimates are sufficiently close. The convergence of this algorithm to the MLEs is established in the Appendix.

%The algorithm proposed above gives the MLEs of parameters $\mu$, $\alpha_i$, $\beta_j$ and $\sigma_{ij}^2$ under $\Omega$.

The LR statistic for testing $H_{0AB}$ against $H_{1AB}$ is evaluated as
\begin{equation}\label{lkr}
\lambda_{AB}=\prod_{i=1}^a\prod_{j=1}^b\left(\frac{\hat{\sigma}_{ij\Omega}^2}{\hat{\sigma}_{ij\Omega_{0AB}}^2}\right)^{n_{ij}/2}.
\end{equation}
The MLEs $\hat{\sigma}_{ij\Omega_{0AB}}^2$ are the points of convergence of  $\sigma_{ij\Omega_{0AB}}^{2(m)}$ of Algorithm \ref{algo2}. The LRT rejects $H_{0AB}$ against $H_{1AB}$ at level of significance $\alpha$ if $\lambda_{AB}<c_{AB}$, where $c_{AB}$ can be found from size condition $\sup_{H_{0AB}}\left(\lambda_{AB}<c_{AB}\right)=\alpha$. Since $\lambda_{AB}$ is not available in a closed analytical form, we cannot get the null distribution $\lambda_{AB}.$

Wilk's Theorem states that $-2\log \lambda_{AB}$ converges in distribution to $\chi_{(a-1)(b-1)}^2$ under $H_{0AB}$.  Therefore, the asymptotic LRT rejects $H_{0AB}$ at level of significance $\alpha$ if $-2\log\lambda_{AB}>\chi_{(a-1)(b-1),\alpha}^2$, where $\chi_{f,\alpha}^2$ indicates the upper $100(1-\alpha)\%$ percentage point of the Chi-square distribution with degrees of freedom $f$.

\subsection{Parametric bootstrap LRT}\label{sec2.3}

As we saw in the previous sections, there are no closed analytical forms for the MLEs of parameters under $\Omega_{0AB}$. The closed analytic form of $\lambda_{AB}$ is thus impossible to obtain. It is also anticipated that limited sample sizes would result in poor performance from the asymptotic LRT. Therefore, when sample sizes are limited, the critical value, say,  $c_{AB}^*$, is found using the parametric bootstrap approach. If the observed value of $\lambda_{AB}$ is less than $c_{AB}^*$, the null hypothesis is rejected. This is dependent on sample sizes and sample variances. Here are the specific steps for evaluating bootstrap critical points.
\begin{Algorithm}\label{algo3}
\hspace{2cm}
\begin{itemize}
\item[1.] Determine sample variances using the samples $Y_{ijk}$: $\frac{n_{ij}}{n_{ij}-1}S_{ij}^2=\frac{1}{n_{ij}-1}\sum\limits_{k=1}^{n_{ij}}(Y_{ijk}-\bar{Y}_{ij})^2$. Since, $\lambda_{AB}$ is independent of location parameters under $H_{0AB}$, generate bootstrap samples $Y_{ijk}^*$ from $N\left(0,\frac{n_{ij}}{n_{ij}-1}S_{ij}^2\right)$ for $k=1,\ldots ,n_{ij};i=1,\ldots ,a;j=1,\ldots ,b.$
\item[2.] Now compute bootstrap MLEs of $\sigma_{ij}^2$ under $\Omega$ and $\Omega_{0AB}$, say, $\hat{\sigma}_{ij\Omega}^{2*}$ and $\hat{\sigma}_{ij\Omega_{0AB}}^{2*}$. The MLEs $\hat{\sigma}_{ij\Omega_{0AB}}^{2*}$ can be calculated using Algorithm \ref{algo2}. Then, the bootstrap likelihood ratio test statistic, say $\lambda_{AB}^*$, is 
\begin{equation}\label{bLRT}
\lambda_{AB}^*=\prod_{i=1}^a\prod_{j=1}^b \left(\frac{\hat{\sigma}_{ij\Omega}^{2*}}{\hat{\sigma}_{ij\Omega_{0AB}}^{2*}}\right)^{n_{ij}/2}.    
\end{equation}
\item[3.] Take the $\alpha$-quantile $\lambda_{AB \left([\alpha H]\right)}^*$ as the bootstrap critical point $c_{AB}^*$ after a large number of repetitions of steps 1 and 2 (let's assume $H=5000$).
\end{itemize}
\end{Algorithm}
%The bootstrap critical value $c_{AB}^*$ is the $\alpha$-quantile of the conditional distribution of $\lambda_{AB}^*$ given the observations $Y_{ijk}.$

We demonstrate the following result for the validity of the bootstrap statistic $\lambda_{AB}^*$.  The Appendix presents the proof.
\begin{theorem}\label{pblrt_cov}
Suppose $F_{\lambda_{AB}|H_{0AB}}(x)$ and $F_{\lambda_{AB}^*|\bm{Y}}(x)$ are the distribution of the statistic $\lambda_{AB}$ under $H_{0AB}$ and the conditional distribution of the bootstrap statistic $\lambda_{AB}^*$ given the observations $\bm{Y}$. Then as $\min_{i,j}n_{ij}\to \infty$, $\sup_x\mid F_{\lambda_{AB}|H_{0AB}}(x)-F_{\lambda_{AB}^*|\bm{Y}}(x)\mid\xrightarrow{P}0.$ 
\end{theorem}
%It can be shown that the test statistics $\lambda_{AB}$ and $\lambda_{AB}^*$ are continuous functions of the sample means and variances. This theorem can then be proved by following steps similar to those in Theorem 4 of Mondal et al. \cite{mondal2023testing}.

\subsection{Multiple comparison test for interaction effects}\label{sec2.4}

The LRT is proposed for interaction effects. However, this test is not useful for simultaneous pair-wise comparisons. Therefore, in this section, we give another test. This is helpful to detect which pairs of effects cause the null hypothesis to be rejected.
Define 
\begin{equation}
Q_{ij_1j_2}=\frac{\bar{Y}_{ij_1}-\bar{Y}_{ij_2}-\bar{Y}_{.j_1}+\bar{Y}_{.j_2}}{S_{ij_1j_2}},
\end{equation}
where $S_{ij_1j_2}^2=\left(1-\frac{2}{a}\right)\frac{S_{ij_1}^2}{n_{ij_1}}+\left(1-\frac{2}{a}\right)\frac{S_{ij_2}^2}{n_{ij_2}}+\frac{1}{a^2}\sum\limits_{i=1}^a \left(\frac{S_{ij_1}^2}{n_{ij_1}}+\frac{S_{ij_2}^2}{n_{ij_2}}\right)$ for $1\leq i\leq a;$ $1\leq j_1<j_2\leq b.$
Now, the simultaneous test for $H_{0AB}$ is given by 
\begin{equation}
Q=\max_{1\leq i\leq a;1\leq j_1<j_2\leq b} \mid Q_{ij_1j_2}\mid .
\end{equation}
This test rejects the null hypothesis $H_{0AB}$ if $Q>q_{\alpha\beta}$, where the critical value $q_{\alpha\beta}$ satisfies $\sup_{H_{0AB}}P(Q>q_{\alpha\beta})=\alpha.$ One cannot obtain a nice form of the exact null distribution of $Q$.  Consequently, we employ the parametric bootstrap method. The procedure is similar to Algorithm \ref{algo3}. Step 1 will remain the same. In Step 2, evaluate the statistic $Q_{ij_1j_2}$ using the bootstrap versions of the means and variances, i.e., $Q_{ij_1j_2}^*=\frac{\bar{Y}_{ij_1}^*-\bar{Y}_{ij_2}^*-\bar{Y}_{.j_1}^*+\bar{Y}_{.j_2}^*}{S_{ij_1j_2}^{*2}}.$ Then evaluate the bootstrap version of the test statistic as $Q^*=\max\left\lbrace Q_{ij_1j_2}^*:~1\leq i\leq a;1\leq j_1<j_2\leq b \right\rbrace .$
Now, in Step 3, take the $(1-\alpha)$-quantile of $H$ statistics $Q^*$, say, $q_{\alpha\beta}^*=Q_{\left([(1-\alpha)H]\right)}^*$ as the bootstrap critical value. 

The $100(1-\alpha)\%$ simultaneous confidence interval for $\left(\alpha\beta\right)_{ij_1}-\left(\alpha\beta\right)_{ij_2};~1\leq i\leq a;1\leq j_1<j_2\leq b$ obtained from the above test is $\bar{Y}_{ij_1}-\bar{Y}_{ij_2}-\bar{Y}_{.j_1}+\bar{Y}_{.j_2}\pm q_{\alpha\beta}^* S_{ij_1j_2}$.

The following theorem is proved in the Appendix to demonstrate the validity of the bootstrap statistic $Q^*$.
\begin{theorem}\label{thm4}
Suppose $F_{Q|H_{0AB}}(x)$ and $F_{Q^*|\bm{Y}}(x)$ are respectively the distribution of the statistic $Q$ under $H_{0AB}$ and the conditional distribution of the bootstrap statistic $Q^*$ given the observations $\bm{Y}$. Then as $\min_{i,j}n_{ij}\to \infty$, $\sup_x\mid F_{Q|H_{0AB}}(x)-F_{Q^*|\bm{Y}}(x)\mid\xrightarrow{P}0.$ 
\end{theorem}

\section{Testing for Simple Effects}\label{sec3}

When there are interactions between factors, the effect of the $i$-th level of factor $A$ is not reflected in the main effects. In this situation, it is advantageous to test the simple impacts rather than the main effects. The simple effects of factor $A$ are denoted by $\alpha_i+\gamma_{ij}$ for $i=1,\ldots ,a;j=1,\ldots ,b.$ The effect of the $i$-th level of factor $A$ at the $j$-th level of factor $B$ is represented by it.  
See Ananda and Weerahandi \cite{ananda1997two}, Xu et al. \cite{xu2013parametric}, and Mondal et al. \cite{mondal2023btesting}, for further details on testing simple effects.
In this section, we develop the LRT and multiple comparison test for testing
$H_{0A}^*:\alpha_i+\gamma_{ij}=0;~i=1,\ldots ,a;j=1,\ldots ,b$.

%Here, $\alpha_i+\alpha\beta_{ij}$ denotes the effect of $i$-th level of factor $A$ when it is present in the $j$-th level of factor $B$.
%In this section, we develop the LRT for testing $H_{0\alpha}^*$ against $H_{1\alpha}^*$.
The test statistics for $H_{0B}^*:\beta_j+\gamma_{ij}=0;i=1,\ldots ,a;j=1,\ldots ,b$ can be obtained by symmetry.
Define the null parameter space as
$$\Omega_{0A}^*=\left\lbrace\left(\zeta_1,\ldots ,\zeta_b,\sigma_{11}^2,\ldots ,\sigma_{ab}^2\right):\zeta_j\in\mathbb{R},\sigma_{ij}\in\mathbb{R}^{+},~i=1,\ldots ,a;j=1,\ldots ,b\right\rbrace.$$ The MLEs under the full parameter space $\Omega$ are available in closed form, given in Section \ref{sec2.1}.
%In Algorithm \ref{algo2}, the procedure for finding the MLEs of parameters under $\Omega_{0AB}$ is given.
The MLEs under $\Omega_{0A}^*$ can be derived using the following algorithm.
%This is provided using a methodology analogous to that used for $\Omega$ and $\Omega_{0AB}$.

\begin{Algorithm}\label{algo4}
\hspace{2cm}
\begin{itemize}
\item[Step ($0$, $0$):] Select the starting parameter values as $\zeta_{j\Omega_{0A}^*}^{(0)}=\bar{Y}_{.j}$ and 
$\sigma_{ij\Omega_{0A}^*}^{2(0)}=S_{ij}^2$ for $i=1,\ldots ,a;j=1,\ldots ,b.$
\item[Step ($m$, 1):] Define $u_{ij\Omega_{0A}^*}^{(m-1)}=\frac{n_{ij}}{\sigma_{ij\Omega_{0A}^*}^{2(m-1)}}$ for $i=1,\ldots ,a;j=1,\ldots ,b.$
Then  $\zeta_j$ at $m$-th iteration as $$\zeta_{j\Omega_{0A}^*}^{(m)}=\frac{\sum\limits_{i=1}^a u_{ij\Omega_{0A}^*}^{(m-1)}\bar{Y}_{ij}}{\sum\limits_{i=1}^a u_{ij\Omega_{0A}^*}^{(m-1)}},j=1,\ldots ,b.$$
\item[Step ($m$, 2):] Compute $\sigma_{ij}^2$ at $m$-th iteration as
$$\sigma_{ij\Omega_{0A}^*}^{2(m)}=\frac{n_{ij}-1}{n_{ij}}S_{ij}^2+(\bar{Y}_{ij}-\zeta_{j\Omega_{0A}^*}^{(m)})^2;i=1,\ldots ,a;j=1,\ldots ,b.$$
\end{itemize}
\end{Algorithm}
Stop at Step ($m$, 2) if $\max\limits_{1\leq j\leq b}\mid\zeta_{j\Omega_{0A}^*}^{(m-1)}-\zeta_{j\Omega_{0A}^*}^{(m)}\mid\leq \epsilon$. 

The likelihood ratio takes the form
\begin{equation}
\lambda_{A}^*=\prod_{i=1}^a\prod_{j=1}^b\left(\frac{\hat{\sigma}_{ij\Omega}^2}{\hat{\sigma}_{ij\Omega_{0A}^*}^2}\right)
^{n_{ij}/2}.
\end{equation}
%For practical implementation, we use the parametric bootstrap approach. 
Let $c_A^{**}$ be the bootstrap critical value. It can be calculated by following the steps similar to those in Algorithm \ref{algo3}. Then the LRT rejects $H_{0A}^*$ if $\lambda_A^*<c_A^{**}$. 

The asymptotic LRT rejects $H_{0A}^*$ at level $\alpha$ if $-2\log\lambda_A^*>\chi_{(a-1)b,\alpha}^2$. 

The multiple comparison test is recommended as an alternative to the LRT for evaluating simple effects. Let $\nu_{ij}=\alpha_i+\gamma_{ij}$ and $\hat{\nu}_{ij}=\bar{Y}_{ij}-\bar{Y}_{.j}$ for $i=1,\ldots ,a;j=1,\ldots ,b.$ Define 
$$R_{i_1i_2j}=\frac{\hat{\nu}_{i_1j}-\hat{\nu}_{i_2j}}{S_{\nu i_1i_2j}},~1\leq i_1<i_2\leq a;~1\leq j\leq b,$$
where $S_{\nu i_1i_2j}^2=\frac{S_{i_1j}^2}{n_{i_1j}}+\frac{S_{i_2j}^2}{n_{i_2j}}.$ The test statistic for the global null hypothesis $H_{0A}^*$ is now $$R=\max_{1\leq i_1<i_2\leq a;1\leq j\leq b}\mid R_{i_1i_2j} \mid.$$
This test rejects $H_{0A}^*$ if $R>r_{\nu}^*$ such that $\sup_{H_{0A}^*}\left(R>r_{\nu}^*\right)=\alpha.$ 
%For practical implementation, we use the parametric bootstrap.
Let $r_{\nu}^{**}$ denote the bootstrap critical point. This can be calculated similarly to that for the test $Q$, described in Section \ref{sec2.4}.

The $100(1-\alpha)\%$ bootstrap family-wise confidence intervals for $\nu_{i_1j}-\nu_{i_2j}$ are then obtained as $\bar{Y}_{i_1j}-\bar{Y}_{i_2j}\pm r_{\nu}^{**} S_{\nu i_1i_2j}$, $1\leq i_1<i_2\leq a;1\leq j\leq b.$

\section{Testing for Treatment Effects in Additive Model}\label{sec4}

In this section, we consider the problem of testing for the significance of the treatment effects when interaction is not present. That is, we consider the additive model
\begin{equation}
Y_{ijk}=\mu+\alpha_i+\beta_j+\epsilon_{ijk};~k=1,\ldots ,n_{ij};~i=1,\ldots ,a;~j=1,\ldots ,b.
\end{equation}
The hypothesis testing problem for significant treatment effects due to factor $A$ is $H_{0A}$ vs. $H_{1A}$. The LRT and simultaneous testing are suggested here for testing main effects. Significance of the main effects due to factor $B$ can be tested using these by exchanging the role of factors $A$ and $B$ effects.

\subsection{The Likelihood Ratio Test}\label{sec4.1}

In this section, the LRT for testing $H_{0A}$ against $H_{1A}$ is formulated. In this case, the full and null parameter spaces are respectively $\Omega_{0AB}$ and $\Omega_{0A}^*$. The MLEs of parameters under these parameter spaces are calculated from Algorithms \ref{algo2} and \ref{algo4}. 
The test statistic obtained using likelihood ratio approach is 
\begin{equation}\label{lrt}
\lambda_A=\prod_{i=1}^a\prod_{j=1}^b \left(\frac{\hat{\sigma}_{ij\Omega_{0AB}}^2}{\hat{\sigma}_{ij\Omega_{0A}^*}^2}\right)^{n_{ij}/2}.
\end{equation}
Using the null distribution of $\lambda_A$, the critical $c_A$ can be found such that $\sup_{H_{0A}}P\left(\lambda_A<c_A\right)=\alpha$, where $\alpha$ indicates the degree of significance.
%This is the rejection criterion.

Since the maximum likelihood estimators of the model parameters are obtained through iterative numerical algorithms rather than closed-form solutions, the exact null distribution of the likelihood ratio test statistic $\lambda_A$ is not analytically tractable. Consequently, the parametric bootstrap (PB) procedure is employed to approximate the corresponding critical values.

The PB approach is similar to Algorithm \ref{algo3}. Using bootstrap samples $Y_{ijk}^*$, compute the MLEs of $\sigma_{ij}^2$ under $\Omega$ and $\Omega_{0A}^*$ following Algorithms \ref{algo2} and \ref{algo4}, respectively. Let us call the bootstrap MLEs $\hat{\sigma}_{ij\Omega_{0AB}}^{2*}$ and $\hat{\sigma}_{ij\Omega_{0A}^*}^{2*}$. Then compute the likelihood ratio test statistic, $\lambda_A^*$, as $$\lambda_A^*=\prod_{i=1}^a\prod_{j=1}^b \left(\frac{\hat{\sigma}_{ij\Omega_{0AB}}^{2*}}{\hat{\sigma}_{ij\Omega_{0A}^*}^{2*}}\right)^{n_{ij}/2}.$$ All other steps are the same.
%$$\lambda_A^*=\prod_{i=1}^a\prod_{j=1}^b \left(\frac{\hat{\sigma}_{ij\Omega_{0AB}}^{2*}}{\hat{\sigma}_{ij\Omega_{0A}^*}^{2*}}\right)^{n_{ij}/2}.$$
%One can also use the asymptotic distribution of $\lambda$ to get the critical points when sample sizes are large. Using Wilks' theorem, the asymptotic null distribution of $-2\log \lambda_A$ is $\chi_{(a-1)}^2$. The asymptotic LRT rejects $H_{0A}$ at level $\alpha$ if $-2\log\lambda_A>\chi_{(a-1),~\alpha}^2$, where $\chi_{f,\alpha}^2$ indicates the upper $100(1-\alpha)\%$ percentage point of the Chi-square distribution with degrees of freedom $f$. However, it has been seen in simulation studies that the performance of ALRT is not good for small samples. 

Alternatively, when the sample size is large, the critical values can be obtained from the asymptotic distribution of $\lambda_A$. By Wilks’ theorem, the null distribution of $-2\log \lambda_A$ converges to $\chi^2$ with degrees of freedom $(a-1)$. Accordingly, the asymptotic likelihood ratio test (ALRT) rejects $H_{0A}$ at significance level $\alpha$ if $-2\log\lambda_A>\chi_{(a-1),~\alpha}^2$, where $\chi_{f,\alpha}^2$ indicates the upper $100(1-\alpha)\%$ percentage point of the Chi-square distribution with degrees of freedom $f$. Nonetheless, simulation studies indicate that the ALRT performs poorly when the sample size is small.

\subsection{Multiple Comparison Test Procedure}\label{sec4.2}

It is later noted in Section \ref{sec5} through simulation studies that bootstrap LRT behaves as an exact test and also has very good power performance. However, it is computationally complex for implementation. In this section, we propose another test for testing $H_{0A}$ against $H_{1A}$ based on pairwise absolute differences of treatment means. This is also useful for pairwise comparisons of treatment effects if $H_{0A}$ is rejected.

For that, we define
\begin{equation}
T_{ii'}=\frac{b\left( \bar{Y}_{i.}-\bar{Y}_{i'.}\right)}{\sqrt{\sum\limits_{j=1}^b\left(\frac{S_{ij}^2}{n_{ij}}+\frac{S_{i'j}^2}{n_{i'j}}  \right)}};~1 \leq i<i' \leq a.
\end{equation}
The test statistic for testing $H_{0A}$ against $H_{1A}$ is defined by
\begin{equation}
T=\max\left\lbrace\mid T_{ii'} \mid :~1 \leq i<i' \leq a\right\rbrace.
\end{equation}
This test rejects $H_{0A}$ whenever $T>d$, where the critical value $d$ can be calculated from the size-$\alpha$ condition $\sup_{H_{0A}}P\left(T>d\right)=\alpha$.  
The process of determining the bootstrap critical value is identical to that outlined in Algorithm \ref{algo3}.
In Step 2, one computes the bootstrap counterpart of $T$ as
$$T^*=\max \left\lbrace \mid T_{ii'}^*\mid :~1 \leq i<i' \leq a\right\rbrace ,$$
where $$T_{ii'}^*=\frac{b\left( \bar{Y}_{i.}^{*}-\bar{Y}_{i'.}^{*}\right)}{\sqrt{\sum\limits_{j=1}^b \left(\frac{S_{ij}^{2*}}{n_{ij}}+\frac{S_{i'j}^{2*}}{n_{i'j}}\right)}};~1 \leq i<i' \leq a,$$ $\bar{Y}_{ij}^*=\frac{1}{n_{ij}}\sum\limits_{k=1}^{n_{ij}}Y_{ijk},~\bar{Y}_{i.}^*=\frac{1}{b}\sum\limits_{j=1}^b\bar{Y}_{ij}^*,S_{ij}^{2*}=\frac{1}{n_{ij}-1}\sum\limits_{k=1}^{n_{ij}}\left(Y_{ijk}^*-\bar{Y}_{ij}^*\right)^2$ for $i=1,\ldots ,a;j=1,\ldots ,b.$

Here, the null hypothesis $H_{0A}$ is rejected if the observed value of the test statistic is more than the critical point. So, the bootstrap critical value is set to the $(1-\alpha)$-quantile of the conditional distribution of $T^*$ based on observations. Therefore, in Step 3, $T_{[(1-\alpha)H]}^*$ has to be chosen as the bootstrap critical value. 

In this case, $H_{0A}$ is rejected when the observed test statistic exceeds the critical value. Accordingly, the bootstrap critical value is taken as the 
$(1-\alpha)$-quantile of the conditional distribution of $T^*$ given the data. Thus, in Step 3, 
 $T_{[(1-\alpha)H]}^*$ is selected as the bootstrap critical value.

For pairwise comparisons, we test $$H_{0Aii'}:\alpha_i=\alpha_{i'}~\text{against}~ H_{1Aii'}:\alpha_i\neq \alpha_{i'}~\text{for all pairs}~ i,i'=1,\ldots ,a;i\neq i'.$$ The null hypothesis $H_{0Aii'}$ is rejected at significance level $\alpha$ whenever $\mid T_{ii'} \mid>d$. The critical value $d$ may be chosen either as the bootstrap critical value $d^*$ or as the asymptotic critical value $Z_{1-\alpha}$ (see Section \ref{sec4.3}).  

Based on the bootstrap critical value, the $100(1-\alpha)\%$ simultaneous confidence interval for $\alpha_i-\alpha_{i'},1\leq i<i'\leq a$ is given by $\bar{Y}_{i.}-\bar{Y}_{i'.}\pm \frac{T_{[(1-\alpha)H]}^{*}}{b}\sqrt{\sum\limits_{j=1}^b \left(\frac{S_{ij}^2}{n_{ij}}+\frac{S_{i'j}^2}{n_{i'j}}\right)}$.  

\subsection{Asymptotic tests}\label{sec4.3}

This section derives the null distribution of $\bm{T}=\left(T_{12},\ldots,T_{1a},T_{23},\ldots,T_{2a},\ldots,T_{(a-1),a}\right)^T$ under the asymptotic condition
\begin{equation}\label{ac1}
\frac{n_{ij}}{N}\to r_{ij};~i=1,\ldots ,a;j=1,\ldots ,b
\end{equation}
as $\min\limits_{i,j} n_{ij}\to \infty$, where $N$ represents the total sample size. 
We define $Z_{ii'}=\sqrt{N}\left(\bar{Y}_{i.}-\bar{Y}_{i'.}\right)$ and $\hat{\tau}_{ii'}^2=\frac{N}{b^2}\sum\limits_{j=1}^b \left(\frac{S_{ij}^2}{n_{ij}}+\frac{S_{i'j}^2}{n_{i'j}}\right)$ for $1 \leq i<i' \leq a.$ Let $\tau_{ii'}^2=\frac{1}{b^2}\sum\limits_{j=1}^b \left(\frac{\sigma_{ij}^2}{r_{ij}}+\frac{\sigma_{i'j}^2}{r_{i'j}}\right)$ for $1 \leq i<i' \leq a$ and $\eta_i^2=\frac{1}{b^2}\sum\limits_{j=1}^b\frac{\sigma_{ij}^2}{r_{ij}}$ for $i=1,\ldots ,a$. 
Note that the distribution of $\left(\bar{Y}_{1.},\ldots ,\bar{Y}_{a.}\right)\sim \bm{N}_a\left(\bm{\mu},\bm{D}_y\right),$ where $\bm{\mu}=\left(\mu,\ldots ,\mu\right)$, $\bm{D}_y$ is a diagonal matrix with diagonal entices $d_{iy}=\frac{1}{b^2}\sum\limits_{j=1}^b\sigma_{ij}^2$ for $i=1,\ldots ,a$. 
Under the asymptotic conditions (\ref{ac1}), 
$\hat{\tau}_{ii'}^2\to \tau_{ii'}^2$ for $1 \leq i<i' \leq a.$ 
Under $H_{0A}$ and the conditions (\ref{ac1}),
Var$(Z_{ii'})\to\tau_{ii'}^2$, Cov$(Z_{ii'},Z_{i'l})\to-\eta_j^2$, Cov$(Z_{ii'},Z_{il})\to\eta_i^2$ and Cov$(Z_{il},Z_{i'l})\to\eta_l^2$ for $1\leq i<i'\leq a$ and $1\leq i<l\leq a.$ 
In all other cases, the covariance is zero.
%$$Z_{ii'}\sim N(0,\rho_{ii'}^2);~1 \leq i<i' \leq a.$$
Under the asymptotic conditions (\ref{ac1}), the covariance matrix of $\bm{Z}=\left(Z_{12},\ldots ,Z_{1a},Z_{23},\ldots ,Z_{2a},\ldots ,Z_{(a-1),a}\right)^T$ converges to $\bm{\Sigma}_z$, where $\bm{\Sigma}_z$ is a partitioned matrix of order $q, ~q=\frac{a(a-1)}{2}.$ The matrix is given by
\[\bm{\Sigma}_z =
\begin{bmatrix}
\vspace{0.1cm}
\bm{M}_{a-1}^1 & \bm{M}_{a-2}^1 & \bm{M}_{a-3}^1 & \bm{M}_{a-4}^1 & \bm{M}_{a-5}^1 & \ldots & \bm{M}_1^1\\
\vspace{0.1cm}
\bm{M}_{a-1}^2 & \bm{M}_{a-2}^2 & \bm{M}_{a-3}^2 & \bm{M}_{a-4}^2 & \bm{M}_{a-5}^2 & \ldots & \bm{M}_1^2\\
\vspace{0.1cm}
\bm{M}_{a-1}^3 & \bm{M}_{a-2}^3 & \bm{M}_{a-3}^3 & \bm{M}_{a-4}^3 & \bm{M}_{a-5}^3 & \ldots & \bm{M}_1^3\\
\vspace{0.1cm}
\bm{M}_{a-1}^4 & \bm{M}_{a-2}^4 & \bm{M}_{a-3}^4 & \bm{M}_{a-4}^4 & \bm{M}_{a-5}^4 & \ldots & \bm{M}_1^4\\
\ldots & \ldots & \ldots & \ldots & \ldots & \ldots &\ldots \\
\bm{M}_{a-1}^{a-1} & \bm{M}_{a-2}^{a-1} & \bm{M}_{a-3}^{a-1} & \bm{M}_{a-4}^{a-1} & \bm{M}_{a-5}^{a-1} &\ldots & \bm{M}_1^{a-1}
\end{bmatrix}.
\]
The diagonal blocks of the above partitioned matrix are:
$\bm{M}_{a-1}^1=\text{diag}\left(\eta_2^2,\eta_3^2,\ldots ,\eta_a^2\right)+\bm{1}_{a-1}\bm{1}_{a-1}^T\eta_1^2$, 
$\bm{M}_{a-2}^2=\text{diag}\left(\eta_3^2,\eta_4^2,\ldots ,\eta_a^2\right)+\bm{1}_{a-2}\bm{1}_{a-2}^T\eta_2^2$, and so on $\bm{M}_{2}^{a-2}=\text{diag}\left(\eta_{a-1}^2,\eta_{a}^2\right)+\bm{1}_2\bm{1}_2^T\eta_{a-2}^2$, $\bm{M}_1^{a-1}=\eta_{a-1}^2+\eta_{a}^2.$ In general, $\bm{M}_{a-l}^l=\left(\eta_{l+1}^2,\ldots ,\eta_a^2\right)+\bm{1}_{a-l}\bm{1}_{a-l}^T\eta_l^2$ for $l=1,2,\ldots ,(a-1)$, where $\bm{1}_{p}$ is a $p$-dimensional vector containing only 1's. The blocks $\bm{M}_{a-l}^l$ are square matrices and their subscripts denote their orders. 
The blocks above the diagonal are
\[\bm{M}_{a-2}^1=
\begin{bmatrix}
-\eta_2^2 & -\eta_2^2 & \ldots & -\eta_2^2\\
\eta_3^2 & 0 & \ldots & 0\\
0 & \eta_4^2 & \ldots & 0\\
\ldots & \ldots & \ldots & 0\\
0 & 0 & \ldots & \eta_a^2\\
\end{bmatrix},
\bm{M}_{a-3}^1=
\begin{bmatrix}
0 & 0 & \ldots & 0\\
-\eta_3^2 & -\eta_3^2 & \ldots & -\eta_3^2\\
\eta_4^2 & 0 & \ldots & 0\\
0 & \eta_5^2 & \ldots & 0\\
\ldots & \ldots &\ldots &\ldots\\
0 & 0 & \ldots & \eta_a^2\\
\end{bmatrix},
\]
\[
\bm{M}_{a-4}^1=
\begin{bmatrix}
0 & 0 & \ldots & 0\\
0 & 0 & \ldots & 0\\
-\eta_4^2 & -\eta_4^2 & \ldots & -\eta_4^2\\
\eta_5^2 & 0 & \ldots & 0\\
0 & \eta_6^2 &\ldots & 0\\
\ldots & \ldots & \ldots & \ldots \\
0 & 0 & \ldots & \eta_a^2\\
\end{bmatrix}, 
\text{and so on}~
\bm{M}_1^1=
\begin{bmatrix}
0\\
0\\
\cdots\\
-\eta_{a-1}^2\\
\eta_a^2
\end{bmatrix}.
\]
In general, the sub-matrix $\bm{M}_{a-l}^1$ is of order $(a-1)\times (a-l)$  for $l=2,3,\ldots ,(a-1).$ The entries of this sub-matrix are as follows: the first $(l-2)$ rows are all zero, the elements in $(l-1)$-th row are all equal to $-\tau_l^2$ and the rest of the elements form a diagonal matrix with diagonal entries $\left(\eta_{l+1}^2,\eta_{l+2}^2,\ldots ,\eta_{a}^2\right)$ for $l=2,3,\ldots ,(a-1).$
The sub-matrices
$\bm{M}_{a-3}^2,\bm{M}_{a-4}^2,\bm{M}_{a-5}^2,\ldots ,\bm{M}_1^2$ are obtained from $\bm{M}_{a-3}^1,\bm{M}_{a-4}^1,\bm{M}_{a-5}^1,\ldots ,\bm{M}_1^1$ by deleting the 1st row; $\bm{M}_{a-4}^3,\bm{M}_{a-5}^3,\bm{M}_{a-6}^3,\ldots ,\bm{M}_1^3$ are obtained from $\bm{M}_{a-4}^1,\bm{M}_{a-5}^1,\ldots ,\bm{M}_1^1$ by deleting the 1st two rows;
and so on $\bm{M}_1^{a-2}$ is obtained from $\bm{M}_1^1$ after deleting first $(a-3)$ rows. 
In general, the sub-matrix at position $(l,l')$ can be obtained from the sub-matrix at position $(1,l')$ by deleting first $(l-1)$ rows for $2\leq l<l'\leq (a-1).$ 
That is, $\bm{M}_{a-l'}^{l}$ can be obtained from $\bm{M}_{a-l'}^{1}$ by deleting first $(l-1)$ rows for $2\leq l<l'\leq (a-1).$

The submatrices located below the main diagonal are the transposes of the corresponding submatrices above the diagonal. Specifically, for $1 \leq l<l' \leq (a-1)$, the submatrix in position $(l,l')$ is given by $\bm{M}_{a-l}^{l'}=\left(M_{a-l'}^l\right)^T.$

%The below diagonal sub-matrices are the transpose of the above diagonal sub-matrices. The sub-matrix at position $(l',l)$ is the transpose of the sub-matrix at position $(l,l')$ for $1\leq l<l'\leq (a-1)$. That is, $\bm{M}_{a-l}^{l'}=\left(M_{a-l'}^l\right)^T,$ $1 \leq l<l' \leq (a-1)$. 

Let $\bm{D}$ be a diagonal matrix whose diagonal elements are $\tau_{12}$, $\tau_{13}$, $\ldots$, $\tau_{1a}$, $\tau_{23}$, $\ldots$, $\tau_{2a}$, $\ldots$, $\tau_{a-2,~a-1}$, $\tau_{a-2,~a}$, $\tau_{a-1,~a}.$
According to the Cramér–Wold theorem, the vector $\bm{Z}$ converges in distribution under the null hypothesis to a multivariate normal distribution,
$\bm{N_{q}}\left(\bm{0},\bm{\Sigma}_z\right)$.
Using the continuous mapping theorem together with the multivariate Slutsky’s theorem, the distribution of $\bm{T}$ under $H_{0A}$ and asymptotic conditions (\ref{ac1}) is given by $\bm{N_{q}}\left(\bm{0},\bm{\Sigma}_T\right),$ where $\bm{\Sigma}_T=\bm{D}^{-1}\bm{\Sigma}_z\bm{D}^{-1}.$

We now introduce an asymptotic test, referred to as the asymptotic MCT. 
From the rejection rule for $\bm{T}$ it follows that
$$P\left(-d\leq T_{ii'}\leq d;1\leq i<i'\leq a\right)=1-\alpha.$$
Here $d$ corresponds to the $(1-\alpha)$ two-sided quantile of multivariate normal distribution $\bm{N_{q}}\left(\bm{0},\bm{\Sigma}_T\right).$

Since $\sigma_{ij}^2$s are unknown, we formulate the asymptotic MCT to reject $H_{0A}$ whenever
$$T>Z_{(1-\alpha)},$$ where $Z_{(1-\alpha)}$ denotes the two-sided $(1-\alpha)$-quantile of the multivariate normal distribution $\bm{N_{q}}\left(\bm{0},\bm{\hat{\Sigma}}_T\right).$ Here $\hat{\bm{\Sigma}}_T$ is obtained from $\bm{\Sigma}_T$ by replacing $\sigma_{ij}^2$ with $\frac{n_{ij}}{n_{ij}-1} S_{ij}^2$ for $i=1,\ldots,a;j=1,\ldots,b.$
In practice, the function `qmvnorm' from the R package `mvtnorm' can be employed to compute the required quantiles of the multivariate normal distribution.

\section{Size and Power Values}\label{sec5}

This section compares the size and power values of the proposed tests to those of Xu {\it{et al}}. (2013) and  (2015). Xu {\it{et al}}. (2013) introduced tests for assessing interaction effects and simple effects in a two-way crossed effects model under heteroscedasticity. They proposed exact tests when variances are known. When dealing with unknown variances, they employed the plug-in test statistic. They used estimators to replace the known variances. To determine the critical points, a parametric bootstrap technique was utilized. Xu {\it{et al}}. (2015) developed analogous tests for testing $H_{0A}$ against $H_{1A}$ in the two-way additive model. We denote $F_{AB}$ and $F_A^*$ the parametric bootstrap tests of Xu {\it{et al}}. (2013) for detecting interaction effects and simple effects, respectively. In addition,  $F_A$ denote the parametric bootstrap test proposed by Xu {\it{et al}}. (2015) for assessing treatment effects in the additive model. A comparison between the proposed LRTs ($\lambda_{AB}$ and $\lambda_A^*$) for testing interaction and simple effects and the parametric bootstrap tests of Xu {\it{et al}}. (2013) ($F_{AB}$ and $F_A^*$) is provided in the Supplementary Information (SI). The simulated size and power results are provided in the SI, with key observations summarized here. Here, the simulation results are reported for the proposed LRT ($\lambda_A$), ALRT, and MCT for testing treatment effects in the additive ANOVA model. For comparison, we also include the simulated results of the tests proposed by Xu {\it{et al}}. (2013 and 2015), namely, $F_A^*$ and $F_A$. Since $F_A^*$ is designed for testing simple effects in a two-way model with interaction, we set the interaction terms to zero in order to make it comparable with the tests for treatment effects in the additive ANOVA model. 
%The model of Xu {\it{et al}}. (2013) is different from the model we have considered here. However, assuming no interaction effects we have done the size and power comparison with the proposed tests.

To evaluate the finite-sample size and power of the proposed tests, a two-phase simulation study is conducted. For given sample sizes $n_{ij}$'s, configurations of parameters $\mu$'s, $\alpha_i$'s, $\beta_j$'s and $\sigma_{ij}^2$'s, first samples $Y_{ijk}$ are generated 10000 times from normal distribution with mean $\mu_{ij}=\mu+\alpha_i+\beta_j$ and variances $\sigma_{ij}^2$ for $k=1,\ldots ,n_{ij};~i=1,\ldots ,a;~j=1,\ldots ,b.$  For each generated sample $Y_{ijk}$, we calculate the observed test statistics along with their corresponding bootstrap critical values using Algorithm \ref{algo3} described earlier.
The empirical size or power is estimated as the proportion of rejections of the null hypothesis over 5000 replications. To determine the critical value, 5000 bootstrap samples are drawn for each generated sample.
For the asymptotic tests, the bootstrap step is not required. For the asymptotic LRT, the critical values are given by the upper $100(1-\alpha)\%$ quantiles of the $\chi^2$ distribution with $(a-1)$ degrees of freedom. The `R' function `qmvnorm' can be used to calculate the critical point of the asymptotic MCT. It corresponds to the two-sided $(1-\alpha)$-quantile of the multivariate normal distribution, which can be obtained once the covariance matrix $\bm{\hat{\Sigma}}_T$ is established. For size evaluation, the parameters are chosen from  $\Omega_{0A}$, while for power evaluation, they are taken from the full parameter space $\Omega_{0AB}.$

We present the size and power values of LRT, MCT, asymptotic LRT and asymptotic MCT for testing treatment effects in the two-way additive model discussed in Section \ref{sec4}. We also present the values of size and power of the test of Xu {\it{et al}}. (2015) ($F_A$) and Xu {\it{et al}}. (2015) ($F_A^*$). The setups used for computing size and power are presented in Tables \ref{tab0} and \ref{tab01}, respectively. The size estimates are reported for the cases 
$a=2,b=3$ and $a=6,b=3$. In Table \ref{tab1}, the size values of LRT and the MCT are presented for $a=2$, $b=3$ with Configuration 1. 
In this table, the sample sizes are taken as small. For these sample sizes, the asymptotic tests do not yield the nominal level. Table \ref{tab2} shows the size values of LRT, MCT, and the asymptotic tests for moderate and large sample sizes with $a=2$, $b=3$. Here, the sample sizes and variances are taken under Configuration 2.
Table \ref{tab3} reports the size values of the LRT and MCT for small samples in a two-way additive model with $a=6$ and $b=3$. Here, the choices of sample sizes and variances are taken from Configuration 3. 
%With the same variances, the size values of LRT, MCT and the asymptotic tests are tabulated in Table \ref{tab4} for large samples as given in Configuration 4.					
Tables \ref{tab1} and \ref{tab3} use the identical sample sizes and parameter configurations as Table 1 of Xu {\it{et al}}. (2015). For all these calculations, the nominal level is set to be 0.05. The size values with level 0.01 are tabulated in the SI. The size values of the proposed tests, along with those of the classical $F$-test designed under the assumption of homogeneous error variances, are reported in Tab 9 of the SI. The table considers various combinations of sample sizes and variance settings.

\begin{figure}[htb]
    \centering
     \subfigure[]{\includegraphics[width=170pt]{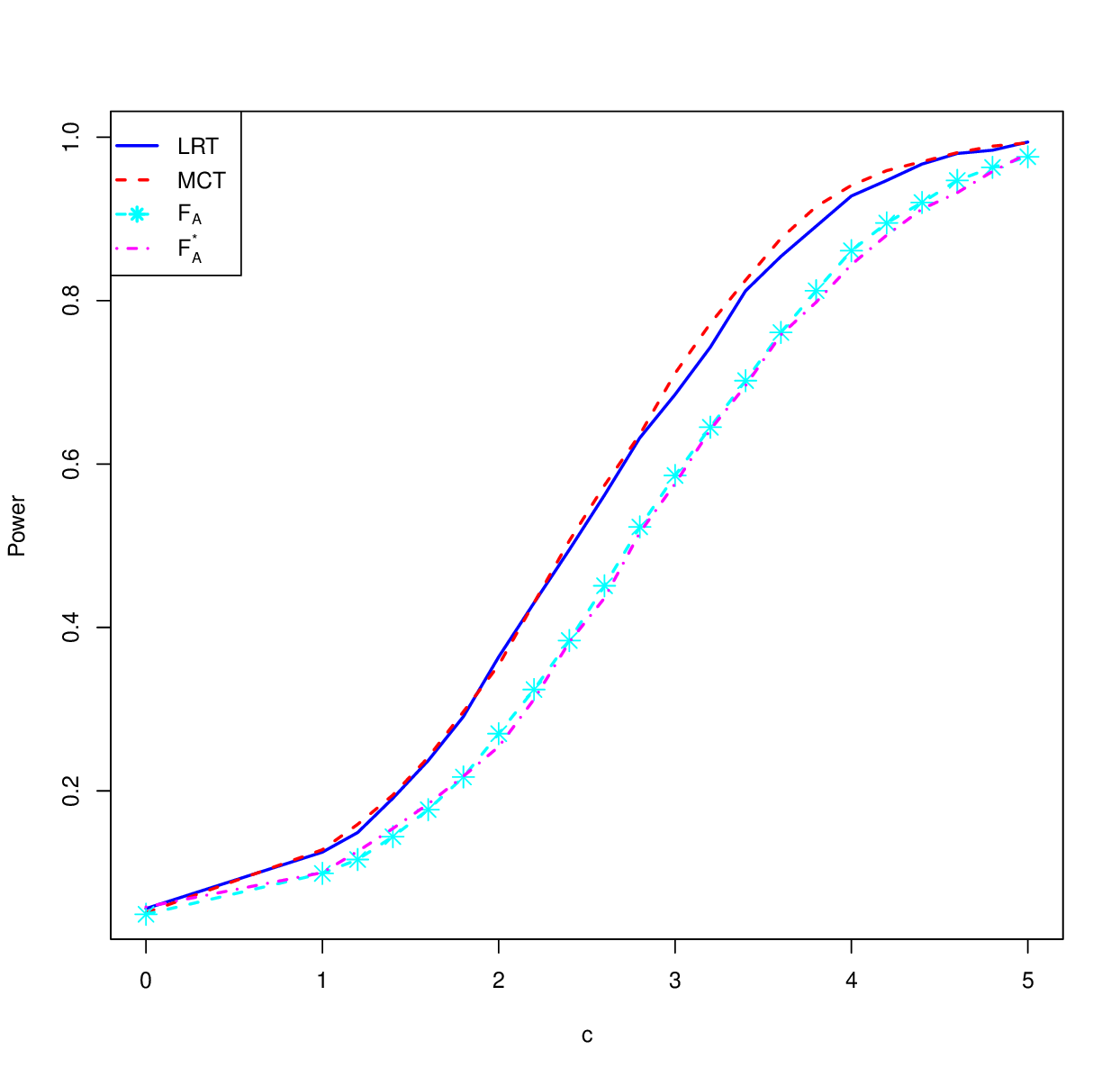}} 
     \subfigure[]{\includegraphics[width=170pt]{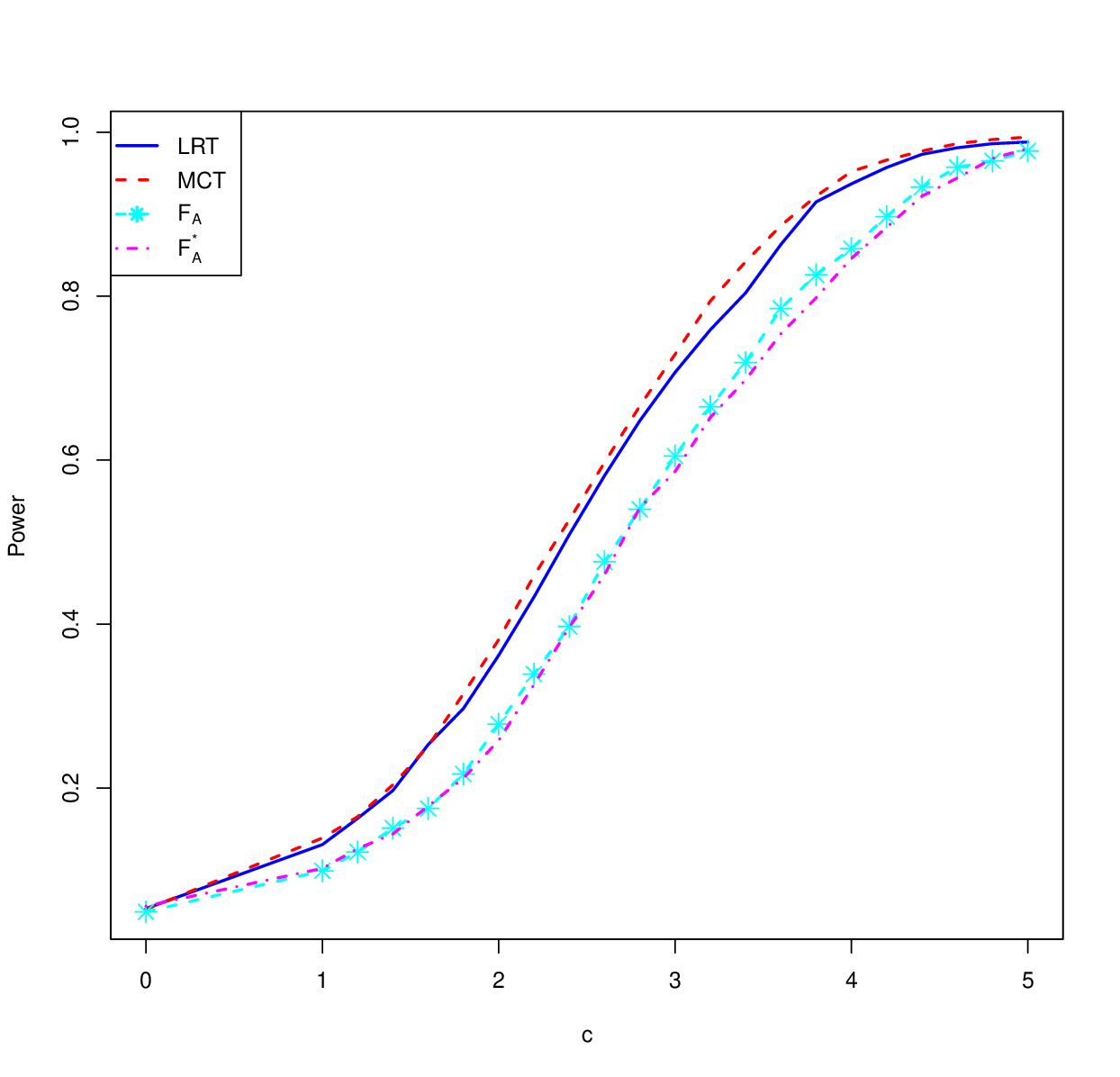}} 
     \subfigure[]{\includegraphics[width=170pt]{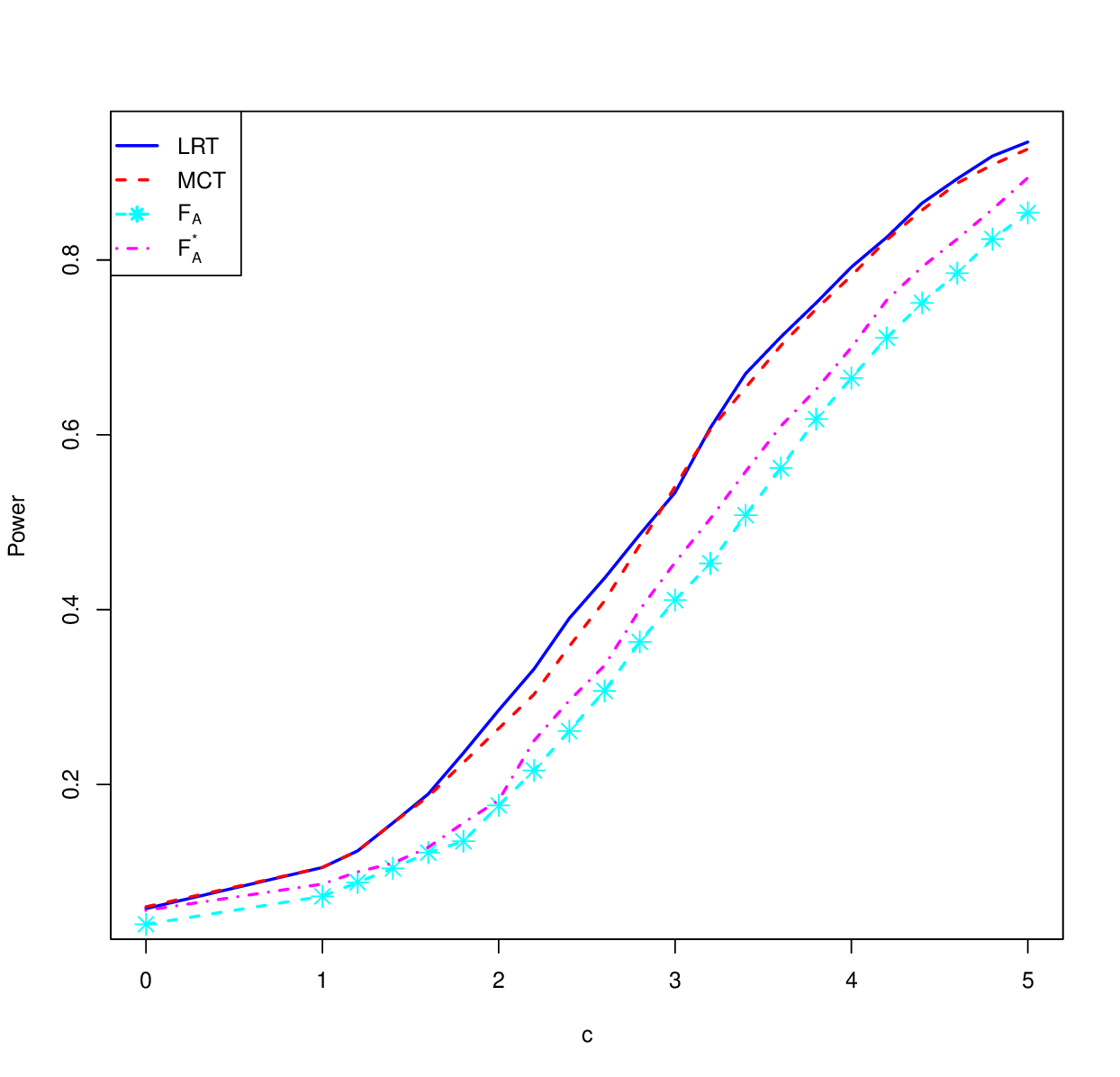}}
     \subfigure[]{\includegraphics[width=170pt]{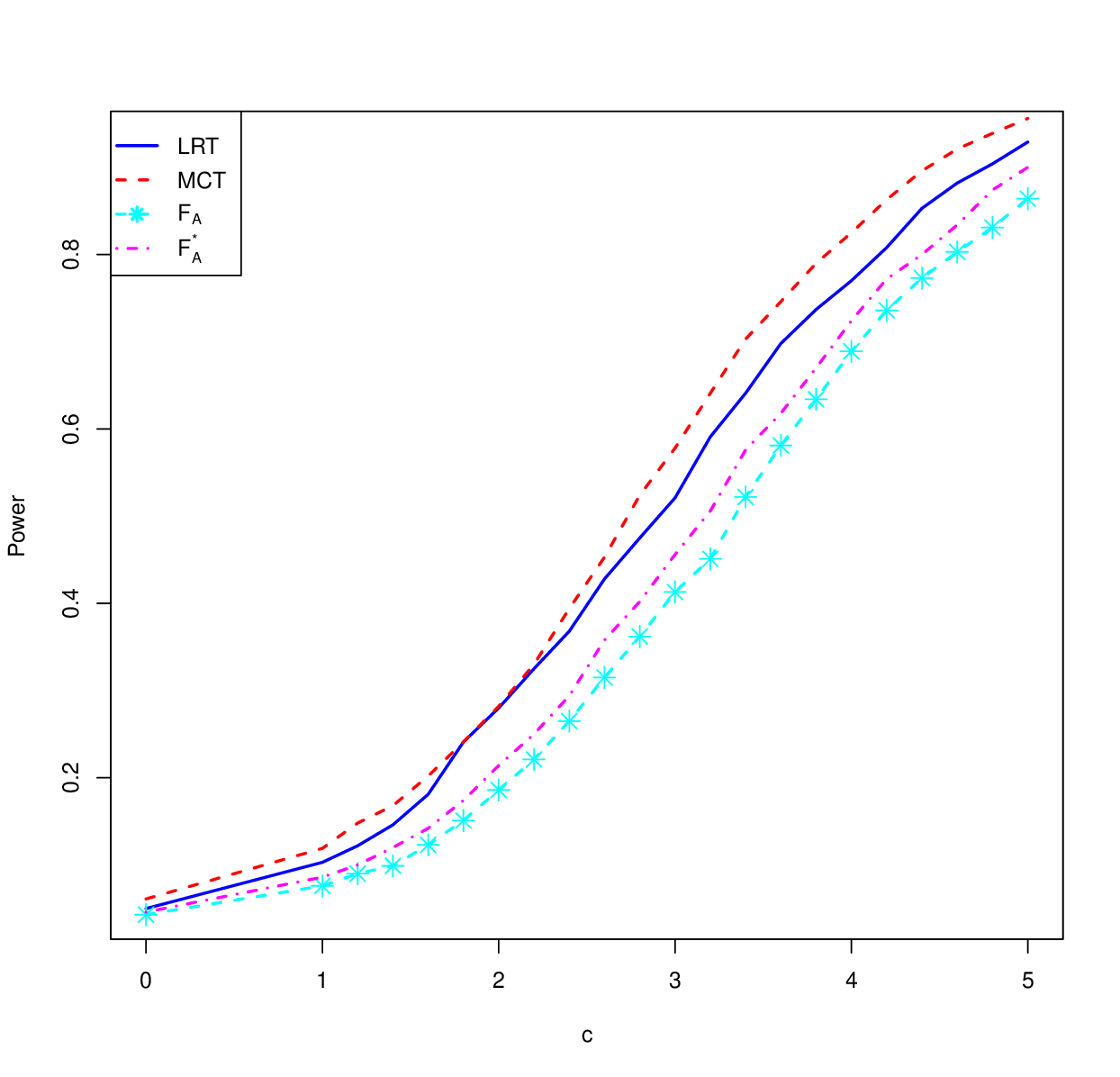}}
   \caption{Power curves for $a=3$, $b=2$, $(\alpha_{1},\alpha_{2},\alpha_{3})$=$c\cdot(0,-0.2,0.2)$, $(\beta_{1}, \beta_{2})$=$(0, 0)$ and sample sizes, variances: (a)   $\left(\mathbf{N_{15}},\boldsymbol{\rho_{12}^{2}}\right)$, (b) $\left(\mathbf{N_{15}},\boldsymbol{\rho_{13}^{2}}\right)$, (c) $\left(\mathbf{N_{16}},\boldsymbol{\rho_{12}^{2}}\right)$ and
   (d) $\left(\mathbf{N_{16}},\boldsymbol{\rho_{13}^{2}}\right)$.}
 \label{fig1}
 \end{figure}

\begin{figure}[htb]
    \centering
     \subfigure[]{\includegraphics[width=170pt]{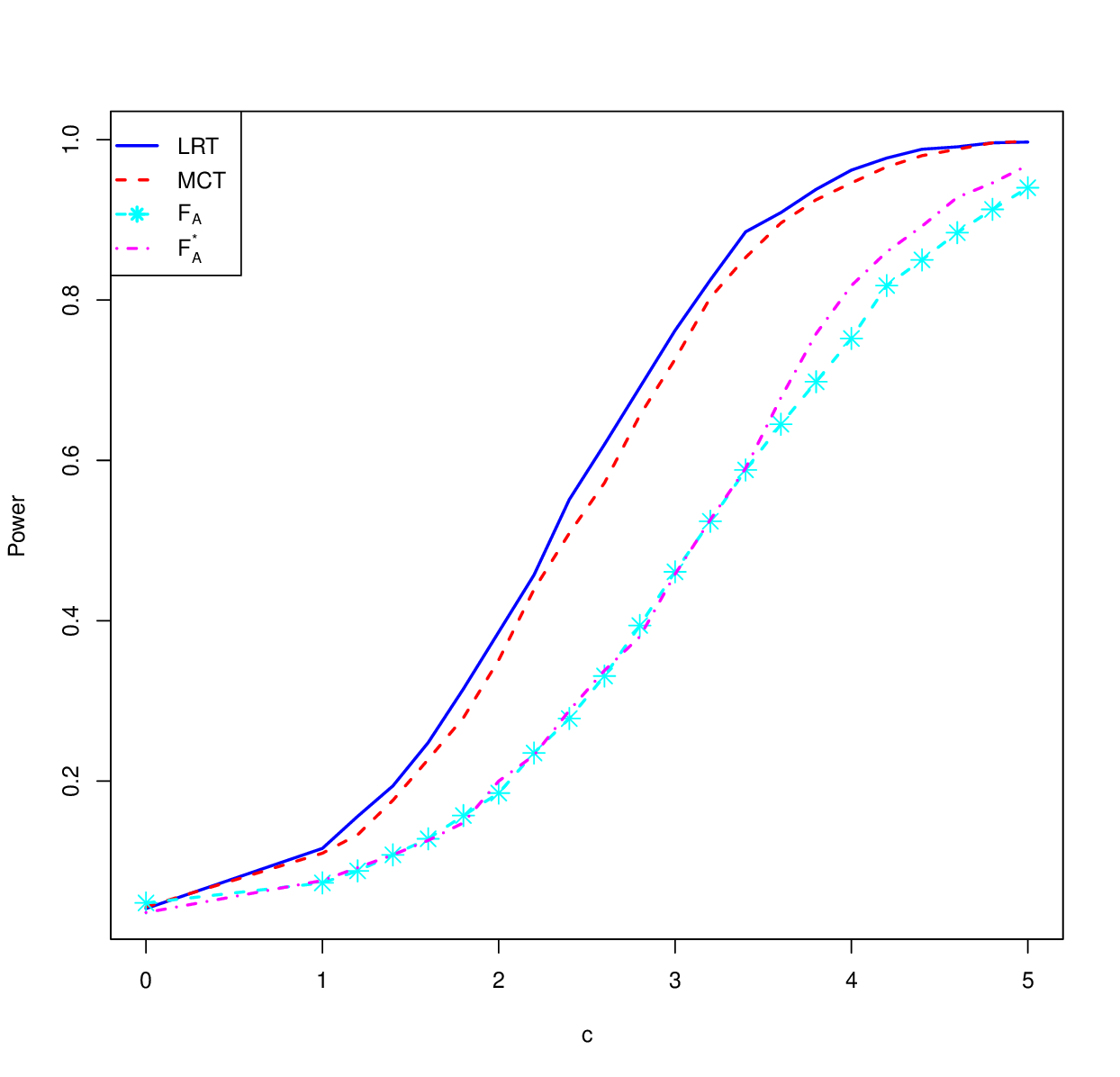}} 
     \subfigure[]{\includegraphics[width=170pt]{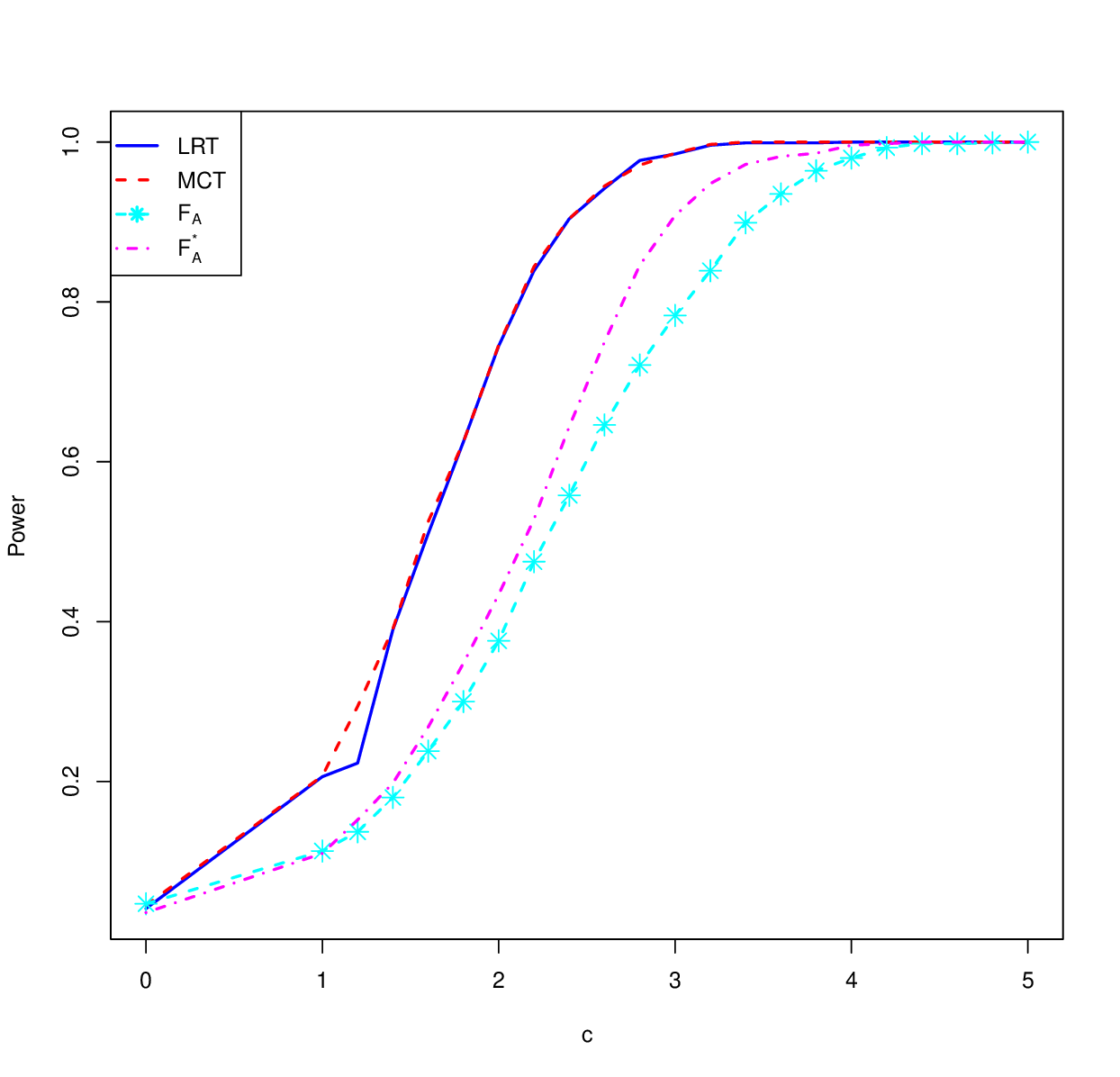}} 
     \subfigure[]{\includegraphics[width=170pt]{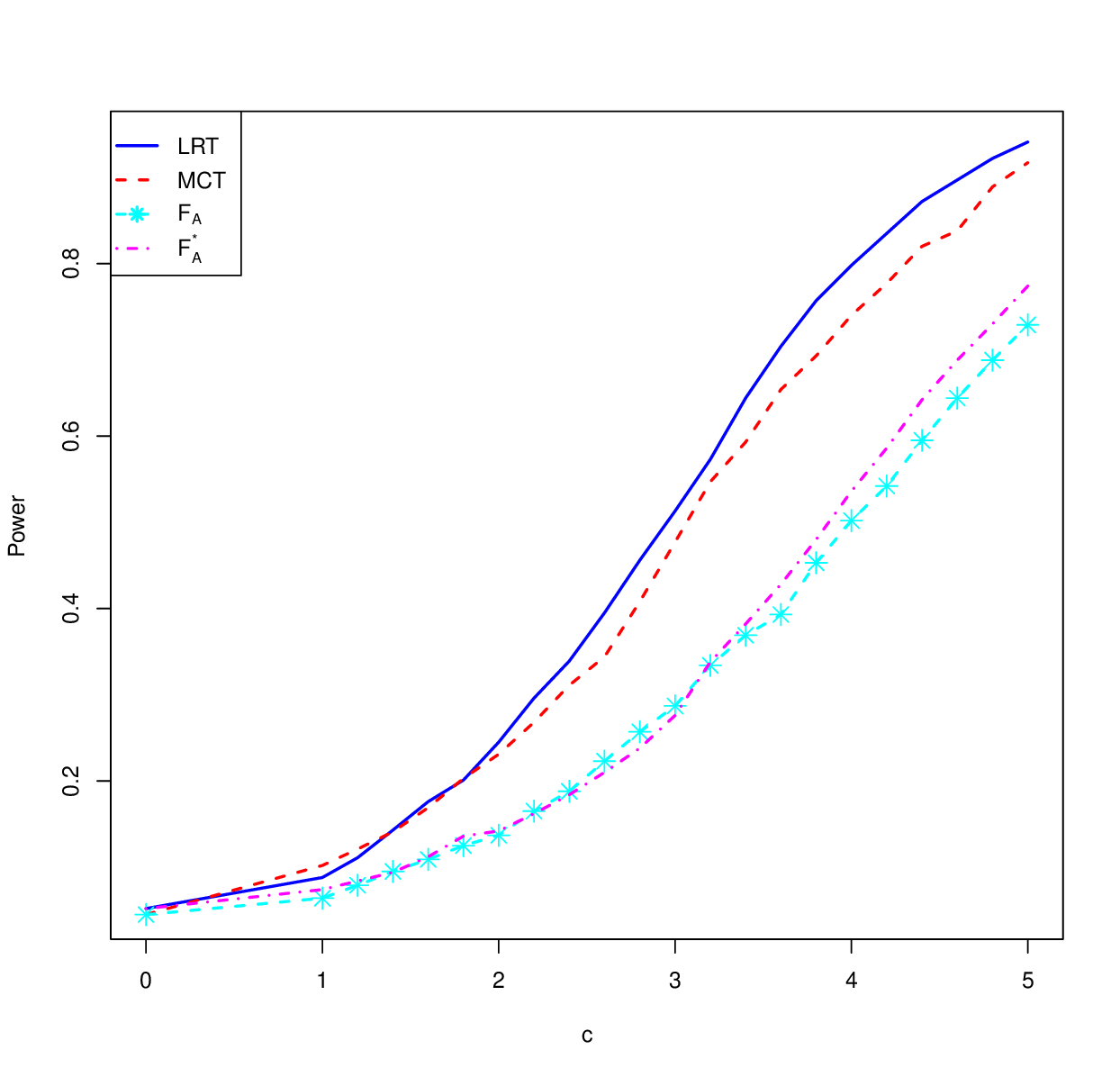}}
     \subfigure[]{\includegraphics[width=170pt]{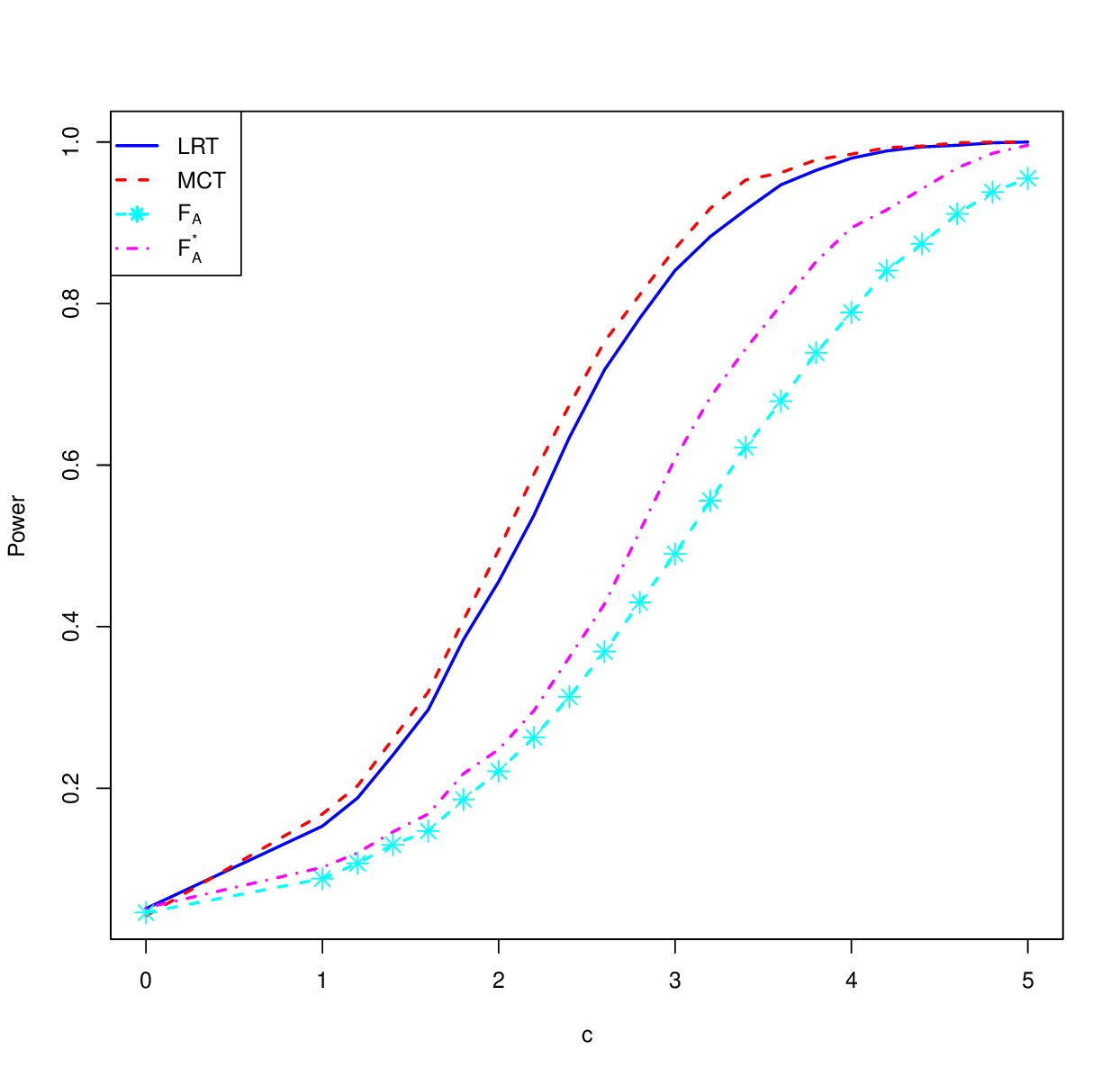}}
   \caption{Power curves for $a=4$, $b=4$, $(\alpha_{1},\alpha_{2},\alpha_{3})$=$c\cdot(1,1.1,1.2,1.3)$, $(\beta_{1}, \beta_{2},\beta_3,\beta_4)$=$(-0.1,0.1,0.2,0.2)$ and sample sizes, variances: (a)   $\left(\mathbf{N_{17}},\boldsymbol{\rho_{14}^{2}}\right)$, (b) $\left(\mathbf{N_{17}},\boldsymbol{\rho_{15}^{2}}\right)$, (c) $\left(\mathbf{N_{18}},\boldsymbol{\rho_{14}^{2}}\right)$ and
   (d) $\left(\mathbf{N_{18}},\boldsymbol{\rho_{15}^{2}}\right)$.}
 \label{fig2}
 \end{figure}

The power plots of the proposed tests are given in Figures \ref{fig1} and \ref{fig2}. Some power values are shown in Tables \ref{tab8} and 
\ref{tab9}. In Figure \ref{fig1}, we take $a=3,b=2$;  $\alpha=c\cdot (0,-0.2,0.2)$; $\beta=(0,0)$ and parameter values are chosen under Configuration 4 of Table \ref{tab01}.
%variances $(\sigma_{11}^2,\sigma_{12}^2,\sigma_{21}^2,\sigma_{22}^2,\sigma_{31}^2,\sigma_{32}^2)$ as $\bm{\rho}_{12}^2$ and $\bm{\rho}_{13}^2$; and sample sizes $(n_{11},n_{12},n_{21},n_{22},n_{31},n_{32})$  as 
%$N_{15}$, $N_{16}$. The choices of parameters are shown in Table \ref{tab01} in Configuration 6. 
In Figure \ref{fig2}, the power values of LRT, MCT and the tests proposed by Xu et al. (2013) and (2015) for $a=4,b=4$ are calculated using the parameters shown at Configuration 5.
%Figure \ref{fig3} presents the power of the LRT, MCT, their asymptotic counterparts, and two existing tests under Configuration 6.	
In Tables \ref{tab8} and \ref{tab9}, the estimated power values are presented for $a=2,b=3$ and $a=6,b=3$, respectively. Here, the sample sizes and parameter combinations are adapted from Xu {\it{et al}}. (2015). %These are shown in  Configurations 9 and 10. 
%In Table \ref{tab8}, the power values of tests are calculated for various values of $\alpha_i$'s and the values of $\beta_j$'s are taken as all zero.

%In the tables of power values and the figures of power curves, $F_A^*$ corresponds to the test of Xu {\it{et al}}. (2013) for testing simple effects with interaction effects, and $F_A$ corresponds to the test of Xu {\it{et al}}. (2015) for treatment effects without interaction effects.
%The power values of Tables \ref{tab8} and \ref{tab9} are calculated using the parameter Configurations 9 and 10, respectively. 

\begin{table}[ht!]
\centering
\caption{Configurations of parameters chosen for size calculations.}
\small
\begin{tabular}{c c c c}
\hline
Configuration & $(a,b)$ & Cell Sizes & Cell Variances \\
\hline

% ---------------- CONFIGURATION 1 ----------------
\multirow{5}{*}{1} 
& \multirow{5}{*}{$(2,3)$}
& $\mathbf{N}_1=(5,5,5,5,5,5)$ 
& $\boldsymbol{\rho}_1^2=(1,1,1,1,1,1)$ \\

& & $\mathbf{N}_2=(10,10,10,10,10,10)$ 
& $\boldsymbol{\rho}_2^2=(0.1,0.1,0.1,0.5,0.5,0.5)$ \\

& & $\mathbf{N}_3=(3,3,4,5,6,6)$ 
& $\boldsymbol{\rho}_3^2=(1,1,1,0.5,0.5,0.5)$ \\

& & $\mathbf{N}_4=(4,6,8,12,16,20)$ 
& $\boldsymbol{\rho}_4^2=(0.1,0.2,0.3,0.4,0.5,1)$ \\

& &  
& $\boldsymbol{\rho}_5^2=(0.3,0.9,0.4,0.7,0.5,1)$ \\
\hline

% ---------------- CONFIGURATION 2 ----------------
2 
& $(2,3)$
& $\mathbf{N}_5=(25,25,25,25,25,25)$ 
& $\boldsymbol{\rho}_1^2$ \\

& & $\mathbf{N}_6=(30,20,25,35,40,30)$ 
& $\boldsymbol{\rho}_2^2$ \\

& & $\mathbf{N}_7=(20,25,30,35,40,45)$ 
& $\boldsymbol{\rho}_3^2$ \\

& & $\mathbf{N}_8=(45,40,35,30,25,20)$ 
& $\boldsymbol{\rho}_4^2$ \\

& &  
& $\boldsymbol{\rho}_5^2$ \\
\hline

% ---------------- CONFIGURATION 3 ----------------
\multirow{7}{*}{3} 
& \multirow{7}{*}{$(6,3)$}
& $\mathbf{N}_9=(5,5,\ldots,5)$ 
& $\boldsymbol{\rho}_6^2=(1,1,\ldots,1)$ \\

& & $\mathbf{N}_{10}=(10,10,\ldots,10)$ 
& \shortstack{
$\boldsymbol{\rho}_7^2=(0.1,0.1,0.2,0.2,0.3,0.3,$\\
$0.4,0.4,0.5,0.5,0.6,0.6,$\\
$0.7,0.7,0.8,0.8,0.9,0.9)$
} \\

& & 
\shortstack{
$\mathbf{N}_{11}=(3,3,3,3,3,4,4,4,4,5,5,5,$\\
$5,6,6,6,6,6)$}
& \shortstack{
$\boldsymbol{\rho}_8^2=(0.1,0.2,0.3,0.4,0.5,$\\
$0.1,0.2,0.3,0.4,0.5,$\\
$0.1,0.2,0.3,0.4,0.5,$\\
$0.2,0.3,2)$
} \\

& & 
\shortstack{
$\mathbf{N}_{12}=(4,4,4,6,6,6,8,8,8,$\\
$12,12,12,16,16,16,20,20,20)$}
& \shortstack{
$\boldsymbol{\rho}_9^2=(0.1,0.1,0.1,\ldots,1)$
} \\

& &
& \shortstack{
$\boldsymbol{\rho}_{10}^2=(0.9,0.8,0.7,0.6,0.5,0.4,$\\
$0.3,0.2,0.1,0.9,0.7,0.6,$\\
$0.5,0.4,0.3,0.2,1)$
} \\

& &
& \shortstack{
$\boldsymbol{\rho}_{11}^2=(0.01,0.01,0.01,0.05,0.05,0.05,$\\
$0.1,0.1,0.1,0.5,0.5,0.5,$\\
$0.6,0.6,0.6,0.8,0.8,1)$
} \\
\hline

\end{tabular}
\label{tab0}
\end{table}

\begin{table}[ht!]
\caption{Configurations of parameters chosen for power calculations}
%\vskip-1cm
%\hrule
\centering
\smallskip

%\tiny
\begin{tabular}{c c c c c}
\hline
\shortstack{Configuration} &\shortstack{$(a,b)$} & \shortstack{Values of $\alpha_i$'s\\ Values of $\beta_j$'s }&\shortstack{Cell Sizes} & \shortstack{Cell\\Variances}\\
\hline
4 & $(3,2)$ & \shortstack{$c\cdot (0,-0.2,0.2)$\\  $(0,0)$} & $\bm{N}_{15}=(10,12,14,10,12,14)$ & $\bm{\rho}_{12}^2=(1,2,3,1,2,3)$\\
 & & & $\bm{N}_{16}=(6,7,8,9,10,11)$ & $\bm{\rho}_{13}^2=(2,2,2,2,2,2)$\\
\hline
5 & $(4,4)$ & \shortstack{$c\cdot (1,1.1,1.2,1.3)$\\ $(-0.1,0.1,0.2,0.2)$} & \shortstack{$\bm{N}_{17}$=(10,12,14,16,10,12,14,16,\\10,12,14,16,10,12,14,16)} & \shortstack{$\bm{\rho}_{14}^2$=(1,2,3,4,1,2,3,4,\\1,2,3,4,1,2,3,4)}\\
& & & \shortstack{$\bm{N}_{18}$=(8,6,7,9,10,12,14,16,10,\\11,12,13,8,6,7,9)} & $\bm{\rho}_{15}^2=(1,1,\ldots ,1)$\\
\hline
\end{tabular}
%\hrule
\label{tab01}
\end{table}

\begin{table}[ht!]
\caption{Simulated size values of LRT ($\lambda_A$), MCT ($T$) and the test of Xu et al. (2013) ($F_A$) for testing treatment effects under Configuration 1; $a=2$ and $b=3$  }
%\vskip-0.3cm
%\hrule
\centering
\smallskip

\small
\begin{tabular}{c c c c c c c c c c c c c}
\hline
 & \multicolumn{3}{c}{$\bm{N}_1$} & \multicolumn{3}{c}{$\bm{N}_2$} & \multicolumn{3}{c}{$\bm{N}_3$} & \multicolumn{3}{c}{$\bm{N}_4$} \\
\hline
$\bm{\sigma}^2$ & LRT & MCT & $F_A$ & LRT & MCT  & $F_A $ & LRT & MCT  & $F_A $ & LRT & MCT  & $F_A $\\
\hline
$\bm{\rho}_1^2$	&	0.053	&	0.046	&	0.040	&	0.051	&	0.049	&	0.050	&	0.044	&	0.046	&	0.050	&	0.053	&	0.051	&	0.060	\\
$\bm{\rho}_2^2$	&	0.052	&	0.047	&	0.050	&	0.053	&	0.051	&	0.050	&	0.046	&	0.047	&	0.040	&	0.048	&	0.039	&	0.050	\\
$\bm{\rho}_3^2$	&	0.052	&	0.050	&	0.040	&	0.057	&	0.054	&	0.050	&	0.057	&	0.050	&	0.06	&	0.050	&	0.047	&	0.060	\\
$\bm{\rho}_4^2$	&	0.055	&	0.053	&	0.050	&	0.049	&	0.051	&	0.050	&	0.044	&	0.043	&	0.040	&	0.047	&	0.044	&	0.040	\\
$\bm{\rho}_5^2$	&	0.061	&	0.052	&	0.040	&	0.051	&	0.047	&	0.050	&	0.043	&	0.042	&	0.050	&	0.058	&	0.046	&	0.050	\\

\hline
\end{tabular}
%\hrule
\label{tab1}
\end{table}

\begin{table}[ht!]
\caption{Simulated size values of LRT ($\lambda_A$), ALRT, MCT ($T$), AMCT, and the test of Xu et al. (2015) ($F_A$)  for  testing treatment effects under Configuration 2; $a=2$ and $b=3$ }
%\vskip-0.3cm
%\hrule
\centering
\smallskip

%\small
\begin{tabular}{c c c c c c c c c c c}
\hline
 &\multicolumn{5}{c}{$N_5$}&\multicolumn{5}{c}{$N_6$}\\
\hline
$\bm{\sigma}^2$ & LRT & ALRT & MCT & AMCT & $F_A$ & LRT & ALRT & MCT & AMCT & $F_A$\\
\hline
$\bm{\rho}_1^2$	&	0.050	&	0.053	&	0.053	&	0.054 & 0.058	&	0.049	&	0.056	&	0.052	&	0.047 & 0.043	\\
$\bm{\rho}_2^2$	&	0.052	&	0.056	&	0.049	&	0.049 & 0.061	&	0.044	&	0.054	&	0.050	&	0.044 & 0.052	\\
$\bm{\rho}_3^2$	&	0.042	&	0.056	&	0.049	&	0.051 & 0.055	&	0.050	&	0.058	&	0.050	&	0.051 & 0.043\\
$\bm{\rho}_4^2$	&	0.051	&	0.058	&	0.046	&	0.048 & 0.061	&	0.052	&	0.056&	0.054	&	0.053	& 0.046\\
$\bm{\rho}_5^2$	&	0.053	&	0.059	&	0.051	&	0.056 & 0.055	&	0.056	&	0.056&	0.054	&	0.053	& 0.045\\
\hline
&\multicolumn{5}{c}{$N_7$}&\multicolumn{5}{c}{$N_8$}\\
\hline
$\bm{\rho}_1^2$	&	0.041	&	0.052	&	0.056	&	0.053 & 0.053	&	0.057	&	0.057	&	0.058	&	0.071 & 0.046\\
$\bm{\rho}_2^2$	&	0.043	&	0.055	&	0.051	&	0.047 & 0.058	&	0.065	&	0.059	&	0.058	&	0.068	& 0.054\\
$\bm{\rho}_3^2$	&	0.039	&	0.054	&	0.051	&	0.049 & 0.051	&	0.055	&	0.058	&	0.058	&	0.068	& 0.049\\
$\bm{\rho}_4^2$	&	0.050	&	0.057	&	0.048	&	0.055 & 0.057 &	0.067	&	0.057	&	0.054	&	0.063& 0.056	\\
$\bm{\rho}_5^2$	&	0.051	&	0.056	&	0.046	&	0.050 & 0.053	&	0.057	&	0.055	&	0.043	&	0.057	& 0.049\\
\hline
\end{tabular}
%\hrule
\label{tab2}
\end{table}

\begin{table}[ht!]
\caption{Simulated size values of LRT ($\lambda_A$), MCT ($T$) and the test of Xu et al. (2015) ($F_A$) under Configuration 3; $a=6$ and $b=3$}

%\vskip-0.3cm
%\hrule
\centering
\smallskip

%\small
\begin{tabular}{c c c c c c c c c c c c c}
\hline
 & \multicolumn{3}{c}{$\bm{N}_{9}$} & \multicolumn{3}{c}{$\bm{N}_{10}$} & \multicolumn{3}{c}{$\bm{N}_{11}$} & \multicolumn{3}{c}{$\bm{N}_{12}$} \\
\hline
$\bm{\sigma}^2$ & LRT & MCT & $F_A$ & LRT & MCT & $F_A$ & LRT  & MCT & $F_A$ & LRT & MCT & $F_A$\\
\hline
$\bm{\rho}_6^2$	&	0.055	&	0.050	&	0.040	&	0.046	&	0.049	&	0.050	&	0.065	&	0.034	&	0.040	&	0.054	&	0.051	&	0.060	\\
$\bm{\rho}_7^2$	&	0.055	&	0.051	&	0.050	&	0.044	&	0.050	&	0.040	&	0.043	&	0.036	&	0.030	&	0.046	&	0.049	&	0.050	\\
$\bm{\rho}_8^2$	&	0.055	&	0.045	&	0.050	&	0.044	&	0.044	&	0.050	&	0.061	&	0.038	&	0.040	&	0.056	&	0.049	&	0.060	\\
$\bm{\rho}_9^2$	&	0.060	&	0.055	&	0.050	&	0.050	&	0.046	&	0.050	&	0.056	&	0.038	&	0.040	&	0.054	&	0.051	&	0.050	\\
$\bm{\rho}_{10}^2$	&	0.066	&	0.053	&	0.050	&	0.045	&	0.045	&	0.050	&	0.066	&	0.039	&	0.040	&	0.047	&	0.047	&	0.060	\\
$\bm{\rho}_{11}^2$	&	0.069	&	0.042	&	0.050	&	0.045	&	0.048	&	0.060	&	0.050	&	0.040	&	0.050	&	0.044	&	0.043	&	0.050	\\

\hline
\end{tabular}
%\hrule
\label{tab3}
\end{table}

\begin{table}[ht!]
\caption{Simulated power values of LRT ($\lambda_A$) and the test proposed by Xu et al. (2013 and (2015) ($F_A^*$ and $F_A$) for testing treatment effects without interactions for $a=2$ and $b=3$, $\bm{\rho}_{16}^2=(0.1,0.1,0.1,0.5,0.5,0.5)$, $\bm{\rho}_{17}^2=(0.3,0.9,0.4,0.7,0.5,1)$, $\bm{N}_{21}=(10,\ldots ,10)$, $\bm{N}_{22}=(10,10,5,5,15,15)$}
%\vskip-0.3cm
%\hrule
\centering
\smallskip

%\small
\begin{tabular}{c c c c c c c c}
\hline
 & \multicolumn{7}{c}{$(\alpha_1,\alpha_2)$}\\
\hline
$(\bm{\sigma}^2,\bm{N})$ & Tests & $(0,0)$ & $(-0.1,0.1)$ & $(0,0.4)$ & $(0,0.6)$ & $(0,0.8)$ & $(0,1)$\\
\hline
$(\bm{\rho}_{16}^2,\bm{N}_{21})$	&	LRT	&	0.053	&	0.252	&	0.732	&	0.968	&	1	&	1	\\
	&	MCT	&	0.051	&	0.268	&	0.778	&	0.985	&	1	&	1	\\
	&	$F_A$	&	0.050	&	0.180	&	0.560	&	0.920	&	1	&	1	\\
    & $F_A^*$ & 0.048 & 0.142 & 0.504  & 0.888  & 0.992  & 1 \\
$(\bm{\rho}_{17}^2,\bm{N}_{21})$	&	LRT	&	0.051	&	0.159	&	0.442	&	0.787	&	0.948	&	0.996	\\
	&	MCT	&	0.047	&	0.161	&	0.488	&	0.814	&	0.964	&	0.999	\\
	&	$F_A$	&	0.050	&	0.100	&	0.310	&	0.620	&	0.880	&	0.990	\\
    & $F_A^*$ & 0.048 & 0.066 & 0.112 & 0.278 & 0.542 & 0.804 \\
$(\bm{\rho}_{16}^2,\bm{N}_{22})$	&	LRT	&	0.048	&	0.256	&	0.739	&	0.983	&	1	&	1	\\
	&	MCT	&	0.052	&	0.253	&	0.679	&	0.961	&	1	&	1	\\
	&	$F_A$	&	0.050	&	0.150	&	0.530	&	0.900	&	0.990	&	1	\\
       & $F_A^*$ & 0.064 & 0.168 & 0.528 & 0.918 & 0.998 & 1 \\
$(\bm{\rho}_{17}^2,\bm{N}_{22})$	&	LRT	&	0.044	&	0.138	&	0.388	&	0.699	&	0.910	&	1	\\
	&	MCT	&	0.054	&	0.140	&	0.420	&	0.728	&	0.917	&	0.980	\\
	&	$F_A$	&	0.050	&	0.090	&	0.250	&	0.510	&	0.728	&	0.990	\\
       & $F_A^*$ & 0.068 & 0.106 & 0.256 & 0.576 & 0.846 & 0.978 \\
\hline
\end{tabular}
%\hrule
\label{tab8}
\end{table}

\begin{table}[ht!]
\caption{Simulated power values of LRT ($\lambda_A$) and the test proposed by Xu et al. (2013 and (2015) ($F_A^*$ and $F_A$) for testing treatment effects without interactions for $a=6$ and $b=3$ under $(\alpha_1,\alpha_2,\alpha_3,\alpha_4)=(0,0,0,0)$, $(\beta_1,\beta_2,\beta_3)=(0,0,0)$, $\bm{N}_{23}=(15,\ldots ,15,20,\ldots ,20,25,\ldots ,25)$, $\bm{N}_{24}=(15,16,17,18,19,20,21,22,23,24,25,26,27,28,29,30,31,32)$, $\bm{\rho}_{18}^2=(0.1,0.1,0.2,0.2,0.3,0.3,0.4,0.4,0.5,0.5,0.6,0.6,0.7,0.7,0.8,0.8,0.9,1)$, $\bm{\rho}_{19}^2=(0.01,0.01,0.01,0.05,0.05,0.05,0.1,0.1,0.1,0.5,0.5,0.6,0.6,0.6,0.8,0.8,1)$ .}
%\vskip-0.3cm
%\hrule
\centering
\smallskip

%\small
\begin{tabular}{c c c c c c c c}
\hline
 & \multicolumn{7}{c}{$(\alpha_5,\alpha_6)$}\\
\hline
$(\bm{\sigma}^2,\bm{N})$ & Tests & $(0,0)$ & $(-0.1,0.1)$ & $(0,0.4)$ & $(0,0.6)$ & $(0,0.8)$ & $(0,1)$\\
\hline

$(\bm{\rho}_{18}^2,\bm{N}_{23})$	&	LRT	&	0.053	&	0.141	&	0.717	&	0.989	&	0.986	&	0.999	\\
	&	MCT	&	0.050	&	0.161	&	0.791	&	0.995	&	1	&	1	\\
	&	$F_A$	&	0.050	&	0.090	&	0.500	&	0.910	&	1	&	1	\\
    & $F_A^*$ & 0.052 & 0.098 & 0.540 & 0.920 & 1 & 1\\
$(\bm{\rho}_{19}^2,\bm{N}_{23})$	&	LRT	&	0.052	&	0.153	&	0.780	&	0.994	&	1	&	1	\\
	&	MCT	&	0.049	&	0.181	&	0.851	&	0.998	&	1	&	1	\\
	&	$F_A$	&	0.050	&	0.100	&	0.620	&	0.970	&	1	&	1	\\
    & $F_A^*$ & 0.038 & 0.100 & 0.582 & 0.950 & 1 & 1\\
$(\bm{\rho}_{18}^2,\bm{N}_{24})$	&	LRT	&	0.046	&	0.140	&	0.818	&	0.998	&	1	&	1	\\
	&	MCT	&	0.037	&	0.164	&	0.888	&	0.999	&	1	&	1	\\
	&	$F_A$	&	0.050	&	0.100	&	0.620	&	0.970	&	1	&	1	\\
    & $F_A^*$ & 0.036 & 0.092 & 0.604  & 0.966  & 1 & 1\\
$(\bm{\rho}_{19}^2,\bm{N}_{24})$	&	LRT	&	0.042	&	0.153	&	0.878	&	0.998	&	1	&	1	\\
	&	MCT	&	0.044	&	0.192	&	0.937	&	1	&	1	&	1	\\
	&	$F_A$	&	0.050	&	0.100	&	0.690	&	0.990	&	1	&	1	\\
& $F_A^*$ & 0.036  & 0.092  & 0.670 & 0.984  & 1 & 1\\

\hline
\end{tabular}
%\hrule
\label{tab9}
\end{table}

From all the tables and figures of size and power values, we have the following observations.	

The observations made from the simulated results for testing treatment effects in an additive model are as follows:
\begin{itemize}
\item[(i)] From Table \ref{tab1}, it is seen that the LRT and MCT behave as exact tests in terms of achieving the nominal level of 0.05. Even for very small samples, these two give exact size values. In Table \ref{tab2}, moderate samples are taken. Here, the asymptotic tests perform as well as the parametric bootstrap tests. Table \ref{tab3} shows that for very small samples, the bootstrap tests achieve the nominal level of 0.05 for $a=6$ and $b=3$.
%Table \ref{tab4} indicates that for large samples the asymptotic tests behave as exact tests. 
Similar observations are made from Tabs. 7 and 8 of the SI with a level of significance of 0.01. 
From Tab. 9, we observe that the classical $F$-test is quite conservative under highly unbalanced samples and heterogeneous variances. However, the proposed tests achieve the nominal level. 

\item[(ii)] Overall, from the numerical results in Tables \ref{tab1} and \ref{tab2}, it is clear that the LRT and MCT based on the parametric bootstrap approach have type I error probabilities quite close to the nominal level.

\item[(iii)] From the estimated power values in Figure \ref{fig1} of the proposed tests and those of the available tests $F_A$ and $F_A^*$, we notice that the proposed LRT and MCT have a substantial power gain over the existing tests. The power values of the LRT and MCT are quite close, and in some situations, MCT gives higher power values than the LRT. 

\item[(iv)] From Figure \ref{fig2} we observe that for all the configurations, the existing tests $F_A$ and $F_A^*$ have smaller power values than those of the LRT and MCT. Here, $F_A^*$ becomes quite conservative. The LRT is more powerful than the MCT when variances are heterogeneous, whereas for equal variances MCT has better performance. 
%In Figure \ref{fig3}, large samples and heterogeneous variances are taken. 
%For large sample sizes, the asymptotic LRT achieves very good power. The existing tests are observed to have the lowest power. Among LRT, MCT, and AMCT, LRT has higher power, and among AMCT and MCT, AMCT has higher power than the MCT. 

\item[(v)] In Tables \ref{tab8} and \ref{tab9}, the existing tests are seen to have the lowest power of all the tests. 

\item[(vi)] We conclude this section by providing an overview of all the simulations carried out for power studies. Overall, it is determined that the PB LRT and MCT perform well irrespective of the number of effects being compared, cell sizes, and cell variance values. With respect to size, the tests effectively control the type I error rate. In terms of power, they consistently outperform the existing tests across all scenarios. Thus, the PB LRT and MCT tests are good alternatives in real-world applications due to their high level of accuracy and ease of implementation. 
\end{itemize}
The observations from the tables of power and size values for testing interaction effects are as follows:
\begin{itemize}
\item[(i)] The simulated size results in Tables 1 and 2 of the SI show that both the LRT and MCT attain the nominal 0.05 significance level for small samples. In some instances with small samples, however, the LRT tends to be conservative. For moderate to large sample sizes, the asymptotic LRT performs well in maintaining the nominal level.
\item[(ii)] From the power values in Tabs 3 and 4 of the SI, it is seen that the asymptotic LRT is superior to all other tests. This is due to the liberal behavior of the asymptotic LRT. The LRT, MCT and the test of Xu et al. (2015) ($F_{AB}$) for testing interaction effects have similar power values. Therefore, when sample sizes are moderate or large, the proposed asymptotic LRT is recommended over all other tests. In the case of finite samples, the parametric bootstrap based LRT and MCT test can be considered as comparable alternatives in practical applications.
\end{itemize}
The observations from the simulated size and power values for testing simple effects are as follows:
\begin{itemize}
\item[(i)] From the simulated size values (Tab 5 of the SI), we notice that the LRT ($\lambda_A^*$) and MCT ($R$) achieve size values close to the nominal level for all configurations of parameters and sample sizes that we have considered here. As expected, ALRT achieves a nominal level of type-II error when sample sizes are moderate and large. When samples are moderate or large, for any configurations of variances, ALRT yields the nominal level of type-I error. 
\item[(ii)] From estimated power values, it is visible that the power values of ALRT are more than those of LRT and the available test $F_A^*$. The power values of LRT, MCT and $F_A^*$ are observed to be similar.
\end{itemize}

It is to be noted that the estimated power value of all the tests increases with the sample size and converges to one, indicating that the proposed procedures are consistent. Moreover, the tests are able to detect small departures from the null hypothesis, indicating satisfactory local power behavior.

\section{Robustness Investigation}\label{sec6}

In this section, the robustness of the proposed test procedures is investigated under the violation of the normality of error distribution.

When the underlying distribution is not normal, the simulation procedures are modified from those mentioned in Section \ref{sec5}. In the outer simulation step, the data is generated using the following strategy: Let $X_{ijk}$ denote a dataset from a particular non normal distribution with mean $\mu_{xij}$ and variance $\sigma_{xij}^2$. For given values of $\mu_{ij}$ and $\sigma_{ij}^2$ we generate $Y_{ijk}$s using the formula $$Y_{ijk}=\mu_{ij}+\sigma_{ij}\frac{X_{ijk}-\mu_{xij}}{\sigma_{xij}};~i=1,\ldots ,a;~j=1,\ldots ,b;~k=1,\ldots ,n_{ij}.$$ The inner bootstrap procedure will remain unchanged. Bootstrap samples have to be generated from the normal distribution only.

We have calculated the size values of LRT and MCT for $a=3,b=2$, given the following underlying distributions: (i) a mixture of N(1,2) and N(2,4) with a mixing parameter of 0.5, (ii) a $t$ distribution with 3 degrees of freedom, (iii) a Weibull distribution with a scale of 1 and a shape of 5, and (iv) a Laplace distribution with a mean of 0 and a scale of 5. The size values of LRT and MCT are presented in Tab 10 and Tab 11 in the SI, respectively. 
%In this instance, we utilize the sample sizes and variances presented in Tab for Configuration 5.

%\newpage

\begin{itemize}
\item[(i)] Table 10 of SI shows that the LRT controls the type-I error for $t_3$, Laplace, and Weibull distribution. Under $t_3$ error scheme with highly unbalanced cell sizes $\bm{N}_{27}$ the LRT is seen to have conservative behaviour. When the mixture of $N(1,2)$ and $N(2,4)$ with mixing proportion 0.5 is considered, the LRT is sometimes liberal and for highly unbalanced sample sizes $\bm{N}_{27}$ it is conservative. 

\item[(ii)] In Table 11 of SI, the  mixture of $N(1,2)$ and $N(2,4)$ with equal probability is considered.  The MCT is seen to achieve the nominal level for slightly unbalanced or equal sample sizes. When the cell sample sizes are moderately or highly unbalanced, the test is sometimes liberal or conservative. Under the $t_3$, Weibull, and Laplace error scheme, the MCT maintains the nominal level.

\item[(iii)] Thus, we may conclude that the LRT and MCT are quite robust for large sample sizes or for less unbalanced cell sizes. 
\end{itemize}
%\newpage
\section{Real Life Example}\label{sec7}
In this section, the applicability of the proposed tests is illustrated using one real data set. 
%For each example, random samples are generated from the original data.
\begin{Example}\label{example1}
In this example, we examine the effects of study time and health status on students’ grades, with particular attention to potential interaction effects. Weekly study time is treated as factor $A$ with four levels: less than 2 hours (ST1), 2–5 hours (ST2), 5–10 hours (ST3), and more than 10 hours (ST4). Health status is considered as factor $B$ with four levels: very bad (scores 1–2), bad (score 3), good (score 4), and very good (score 5). The response variable is the first-period grade (G1). The original data set is available at \url{https://www.kaggle.com/datasets/uciml/student-alcohol-consumption}.From this data, we have extracted the relevant entries. The response variable (G1) is modeled using a two-way ANOVA framework with study time (factor $A$, 4 levels) and health status (factor $B$, 4 levels) as the factors.
Here, $a=4$ and $b=4$. We first examine whether the two factors exhibit significant interaction effects on the response by testing $H_{0AB}:\gamma_{ij}=0~\forall (i,j)$ against its natural alternative $H_{1AB}$. 
Based on the observed values of the test statistics and corresponding critical points (Table \ref{exm1}), neither the LRT nor the MCT rejects the null hypothesis $H_{0AB}$.

In the absence of significant interaction effects, we proceed to test for the main treatment effects of factor $A$, with hypotheses  $H_{0A}:\alpha_i=0~\forall~i$ against its natural alternative $H_{1A}.$
The test statistics and critical values for both LRT and MCT are presented in Table \ref{exm1}. Both tests reject $H_{0A}$, prompting the computation of simultaneous confidence intervals for pairwise differences in treatment effects. These intervals indicate significant differences between the 1st and 3rd levels and between the 2nd and 3rd levels of factor $A$.
\begin{table}[ht!]
\caption{Example 1 results}

%\vskip-0.3cm
%\hrule
\centering
\smallskip

\small
\begin{tabular}{c c c }
\hline
%&\multicolumn{4}{c}{Mixture of N(1,2) and N(2,4)} &\multicolumn{4}{c}{$t_3$}\\
%\hline
Sample Sizes	&	Sample Means	&	Sample Variances	\\
\hline
 \makecell{$n_{11}=20$, $n_{12}=20$, $n_{13}=15$, \\$n_{14}=50$, $n_{21}=56$, $n_{22}=43,$\\ $n_{23}=30,$ $n_{24}=69,$ $n_{31}=9$,\\ $n_{32}=18$, $n_{33}=17$, $n_{34}=21$,\\ $n_{41}=7$, $n_{42}=10$, $n_{43}=4$,\\$n_{44}=6$} &\makecell{$\bar{y}_{11}=11.4$, $\bar{y}_{12}=10,$ $\bar{y}_{13}=10.8,$ \\$\bar{y}_{14}=10.12,$ $\bar{y}_{21}=11.0178,$ $ \bar{y}_{22}=10.1395,$ \\$\bar{y}_{23}=10$, $\bar{y}_{24}=10.9565,$ $\bar{y}_{31}=13.8889$, \\$\bar{y}_{32}=11.8333,$ $\bar{y}_{33}=11.8235,$ $\bar{y}_{34}=11.6191$,\\ $\bar{y}_{41}=12.5714$, $\bar{y}_{42}=11.6$, $\bar{y}_{43}=11.25$, \\$\bar{y}_{44}=12$} & \makecell{$s_{11}^2=13.5158$, $s_{12}^2=10.9474$, $s_{13}^2=19.3143$,\\ $s_{14}^2=11.2098$, $s_{21}^2=10.7088$,  $s_{22}^2=9.5515$,\\ $s_{23}^2=6.5517$, $s_{24}^2=10.8951$, $s_{31}^2=13.3611$,\\
$s_{32}^2=6.9706$, $s_{33}^2=11.5294$, $s_{34}^2=7.3476$,\\ $s_{41}^2=17.6190$, $s_{42}^2=14.0444$, $s_{43}^2=3.5833$,\\ $s_{44}^2=16$)}\\
\hline
\multicolumn{3}{c}{Test Results for Interaction Effects}\\
\hline
Test & Test statistic and critical value & Decision\\
\hline 
LRT & 0.127625 \& $7.92111\times 10^{-05}$ &$H_{0AB}$ is not rejected\\
MCT & 1.1839 \& 3.3304 & $H_{0AB}$ is not rejected\\
\hline
\multicolumn{3}{c}{Test Results for Treatment Effects}\\
\hline
Test & Test statistic and critical value & Decision\\
\hline 
LRT & 0.0003 \& 0.0098 &$H_{0A}$ is rejected\\
MCT &3.6708 \& 2.6471 &$H_{0A}$ is rejected \\
\hline
\multicolumn{3}{c}{Simultaneous Confidence Intervals for Treatment Effects}\\
Difference & Lower Bound & Upper Bound\\
\hline 
$\alpha_1-\alpha_2$ & -1.1909 & 1.2939 \\
$\alpha_1-\alpha_3$ & -3.2821 & -0.1403 \\
$\alpha_1-\alpha_4$ & -3.3908 & 0.8901 \\
$\alpha_2-\alpha_3$ & -3.0339 & -0.4916 \\
$\alpha_2-\alpha_4$ &  -3.2304 & 0.5766 \\
$\alpha_3-\alpha_4$ & -1.6967 & 2.5683 \\
\hline
\end{tabular}
%\hrule
\label{exm1}
\end{table}

\end{Example}

\section{Software Package}\label{sec8}
We have created an R package, \texttt{TwowayANOVATests}, which implements six testing functions: \texttt{LRTInt}, \texttt{LRTsimple}, \texttt{LRTtreat}, \texttt{MCTInt}, \texttt{MCTsimple}, and \texttt{MCTtreat}. The package is publicly available through the GitHub repository \texttt{AnjanaStat} at \url{https://github.com/AnjanaStat/TwowayANOVATests}.
To apply the functions, the user must specify six arguments: the matrix of sample means (mean), sample sizes ($N$), sample variances (var), the number of levels of factor $A$ ($a$), the number of levels of factor $B$ ($b$), and the significance level (alpha). The sample means, sample sizes, and variances should each be supplied as matrices of dimension $a \times b$. Specifically, the entry in the $(i,j)$-th position of these matrices corresponds to the sample mean, sample size, and variance, respectively, for cell $(i,j)$.

\section{Conclusion} 
The hypothesis testing problems in two-way ANOVA models have been studied extensively over the years. However, there are a few studies in which error variances are heterogeneous. Among the available tests, plug-in type tests are known to have the best power performance so far. Due to difficulty in obtaining solutions to the likelihood equations, the likelihood ratio test has not been considered earlier for this problem. In this paper, we have developed several tests considering heterogeneous error variances, namely LRT, MCT, asymptotic LRT, and asymptotic MCT. To facilitate the derivation of the likelihood ratio test statistic, we also provide iterative procedures for computing the maximum likelihood estimators of the parameters in two-way crossed and additive models under heteroscedasticity.
Simulation studies indicate that the proposed LRTs for testing interaction and simple effects exhibit satisfactory size performance. The LRTs achieve power values comparable to those of existing bootstrap tests. Moreover, for moderate to large sample sizes, the asymptotic LRT outperforms all competing methods. In the context of testing treatment effects in the additive ANOVA model, both the LRT and MCT successfully maintain the nominal type-I error rate, even with very small samples and arbitrary variance structures. The asymptotic tests behave well for moderate and large samples. In terms of power, the LRT and MCT are observed to be more powerful than the existing test. The proposed tests have a gain in power of about 30-50$\%$ over the present tests. The robustness of the proposed tests is demonstrated through simulation studies based on empirical size and power values. Furthermore, an R software package has been developed and made publicly available on an open platform to facilitate practical application by experimenters.

\end{document}